\documentclass[useAMS,usenatbib,usegraphicx]{mn2e}

\begin{document}

\title[Diffuse Intracluster Light at Intermediate Redshifts]{Diffuse Intracluster Light at Intermediate Redshifts:\\ICL observations in a X-ray cluster at z=0.29 }

\author[I. Toledo et al.]
  {I.~Toledo,$^1$ J.~Melnick,$^2$ F.~Selman,$^2$ H. Quintana,$^1$ E. Giraud,$^3$ P.~Zelaya,$^1$\\
   $^1$Departamento de Astronom\'ia y Astrof\'isica, Pontificia Universidad Cat\'olica de Chile, Casilla 306, Santiago, Chile\\
   $^2$European Southern Observatory, Alonso de Cordova 3107, Santiago, Chile\\
   $^3$Laboratoire Physique Th\'eorique et Astroparticules (LPTA),\\ Universit\'e Montpellier 2 - CNRS/IN2P3, Place E. Bataillon, 34095 Montpellier, France}

\date{}

\pagerange{\pageref{firstpage}--\pageref{lastpage}} \pubyear{2010}

\maketitle

\label{firstpage}
\begin{abstract}
The diffuse intracluster light (ICL) contains a significant fraction of the total stellar mass in clusters of galaxies, and contributes in roughly equal proportion as the hot intra-cluster medium (ICM) to the total baryon content of clusters. Because of the potential importance of understanding the origin of the ICL in the context of the formation and evolution of structure in the Universe, the field has recently undergone a revival both in the quality and quantity of observational and theoretical investigations. Due to cosmological dimming, the observational work has mostly concentrated on low redshift clusters, but clearly observations at higher redshifts can provide interesting clues about the evolution of the diffuse component. In this paper we present the first results of a program to characterize the ICL of intermediate redshift clusters. We find that at $z\sim0.3$, the X-ray cluster RX J0054.0-2823 already has a significant ICL and that the fraction of the total light in the ICL and the brightest cluster galaxy (BCG) is comparable to that of similar clusters at lower redshift. We also find that the kinematics of the ICL is consistent with it being the remnant of tidally destroyed galaxies streaming in the central regions of the cluster, which has three central giant elliptical galaxies acting as an efficient ``grinding machine''. Our cluster has a bi-modal  radial-velocity distribution and thus two possible values for the velocity dispersion. We find that the cluster fits well in the correlation between BCG+ICL fraction and cluster mass for a range of velocity dispersions, leading us to question the validity of a relevant correlation between these two quantities.
\end{abstract}


\section{Introduction}
\label{SECintro}

\footnotetext[1]{Based on observations collected at the European Organisation for Astronomical Research in the Southern Hemisphere, Chile. ESO 078A--0456; ESO 65.O--0425}

The diffuse intracluster light (ICL) in rich clusters of galaxies is by now a mature field that was pioneered by \cite{Zwicky1951,Zwicky1952} and quantified using innovative photoelectric and photographic techniques by \cite{deVaucouleurs1970}, \cite{Melnick1977} and \cite{Thuan1977}. But it was not until the advent of low-noise panoramic CCD detectors that the field matured and evolved to the present state of the art \citep{Zibetti2005,Gonzalez2007,Mihos2005,Krick2006}. However, even with state of the art CCDs on the most powerful telescopes, the task of detecting and quantifying the diffuse intergalactic component is far from straightforward. Not only are we dealing with signals that are much fainter than the sky, but it is also very difficult to disentangle the emission from bona-fide free floating intergalactic stars from the outer haloes of the brightest cluster galaxies (BCGs). Recently, there have been substantial improvements in the theoretical understanding of the origin of the ICL through detailed numerical simulations \citep[eg.][and references therein]{Rudick2009,Dolag2009a,Puchwein2010,Murante2007}, and the field is evolving from being purely observationally driven, to a phase of rich interaction between theory and observations. 

Numerical simulations indicate that most of the ICL is accrued fairly recently in the merging history of clusters, probably at redshifts $z<1$. Some of the ICL is produced in earlier sub-mergers and  falls into the clusters together with their parent groups and smaller clusters, while much of the intracluster light comes from stars stripped off massive galaxies that fall into the forming cluster.  Recent state of the art hydrodynamical simulations \citep{Puchwein2010} predict that the ICL should contain a significant fraction ($\sim45\%$) of the total stellar mass in clusters, well in excess of the values typically observed. Thus, observations of the diffuse intracluster light in clusters should yield information not only about the origin of the ICL, but also about the formation of the clusters themselves. Mapping the evolution of the ICL as a function of redshift should, at least in principle, provide a novel cosmological probe.  

In the course of an investigation of the diffuse intergalactic light in X-ray emitting clusters at intermediate redshifts, we detected a puzzling S-shaped arc-like structure in the ROSAT cluster RX J0054.0-2823  that we tentatively identified as the gravitationally lensed image of a background galaxy at a redshift between $z=0.5$ and $z=1.0$ \citep{Faure2007}.  The cluster is characterized by  having three dominant D class galaxies, two of which are interacting forming a dumbbell cD system. Thus, although our lensing models reproduced surprisingly well the arc-structure, the possibility that it could be tidal debris related to these massive galaxies could not be excluded.  In fact, numerical simulations of cluster formation show that galaxies and groups falling in the cluster potential actually retain their identities for rather  long times, so the ICL should actually have a filamentary structure, which is actually observed in some clusters \citep[see][ and references therein]{Mihos2005,Rudick2009}. We therefore embarked on an ambitious project to measure the redshift of the arc, which has a low surface brightness so, unless the spectrum is dominated by strong emission lines, required long integration times.  We designed an observing strategy allowing us at the same time to observe the arc, the diffuse intracluster Light (ICL), and a substantial number of individual galaxies in the field taking advantage of the multi-object  spectroscopic mode of the FORS2 instrument on Paranal. This allowed us to obtain the redshifts of more than 650 galaxies in addition to very deep observations of the arc, and very deep images and long-slit observations of the ICL.
 
The analysis of the field population around the cluster are presented in two separate papers \citep{Giraud2010,Giraud2011} that also contain the details of the observations and the data analysis procedures. The present paper is devoted to the investigation of the diffuse intra-cluster light component and the analysis of the dynamics of the cluster based on our deep images and deep spectroscopy.  In Section~\ref{SECobs} we present a detailed description of the observations and the data reduction techniques for the photometry and for the long-slit spectroscopy that has not been described in our previous papers. Section~\ref{results} presents  the dynamical and photometric properties of the cluster and the ICL and details the techniques that we used to detect and measure the ICL component and to define its photometric and spectroscopic properties, including those of the S-shaped arc. Finally, Sections~\ref{SECanal} and \ref{SECconcl} discuss the results and presents the conclusions of this work.  Throughout this work we use the standard cosmology $H_0=75~\mathrm{km~s^{-1}}$, $\Omega_\mathrm{m}=0.3$; $\Omega_\mathrm{\Lambda}=0.7$.

\section{Observations}
\label{SECobs}

\subsection{Photometry}

Our initial deep images of the cluster were obtained with SUSI2 at the La Silla New Technology Telescope (NTT) in the V and I bands in the nights 26 and 27 of September 2000. SUSI2  is an excellent instrument for this kind of work: besides the reflections in the three telescope mirrors it has only one additional reflection, one transmission through the dewar entrance window and the filter, ensuring a very clean point spread function (PSF). The observations were executed using the Continuous Differential Imaging (CDI) method described in detail by \citet{Melnick1999}. Briefly, the observations are done (as in the infrared) by nodding between object and sky. Since SUSI2 has a mosaic of two CCDs, the nodding was done by alternatively centering the brightest cluster galaxy (BCG) on either CCD and thus measuring the sky on the other. The images of each CCD are calibrated using standard techniques, and  the median of the two bracketing sky images is subtracted from each cluster image to obtain a set of sky-subtracted  images of the cluster that we call CDIs. All CDIs are astrometrically registered and photometrically calibrated before being averaged. The resulting image is used to create a mask where every discrete object is flagged. This mask is used to create new CDIs were the signal from discrete objects in the sky frames is removed. A final science image is produced in this way and \citet{Melnick1999} showed that the technique works even under rather unfavorable observing conditions. Table~\ref{tableob} summarizes the number of consecutive images used for each band, and the total amount of observing time.

\begin{figure}
  \hspace*{-0.5cm}\includegraphics[width=7.5cm]{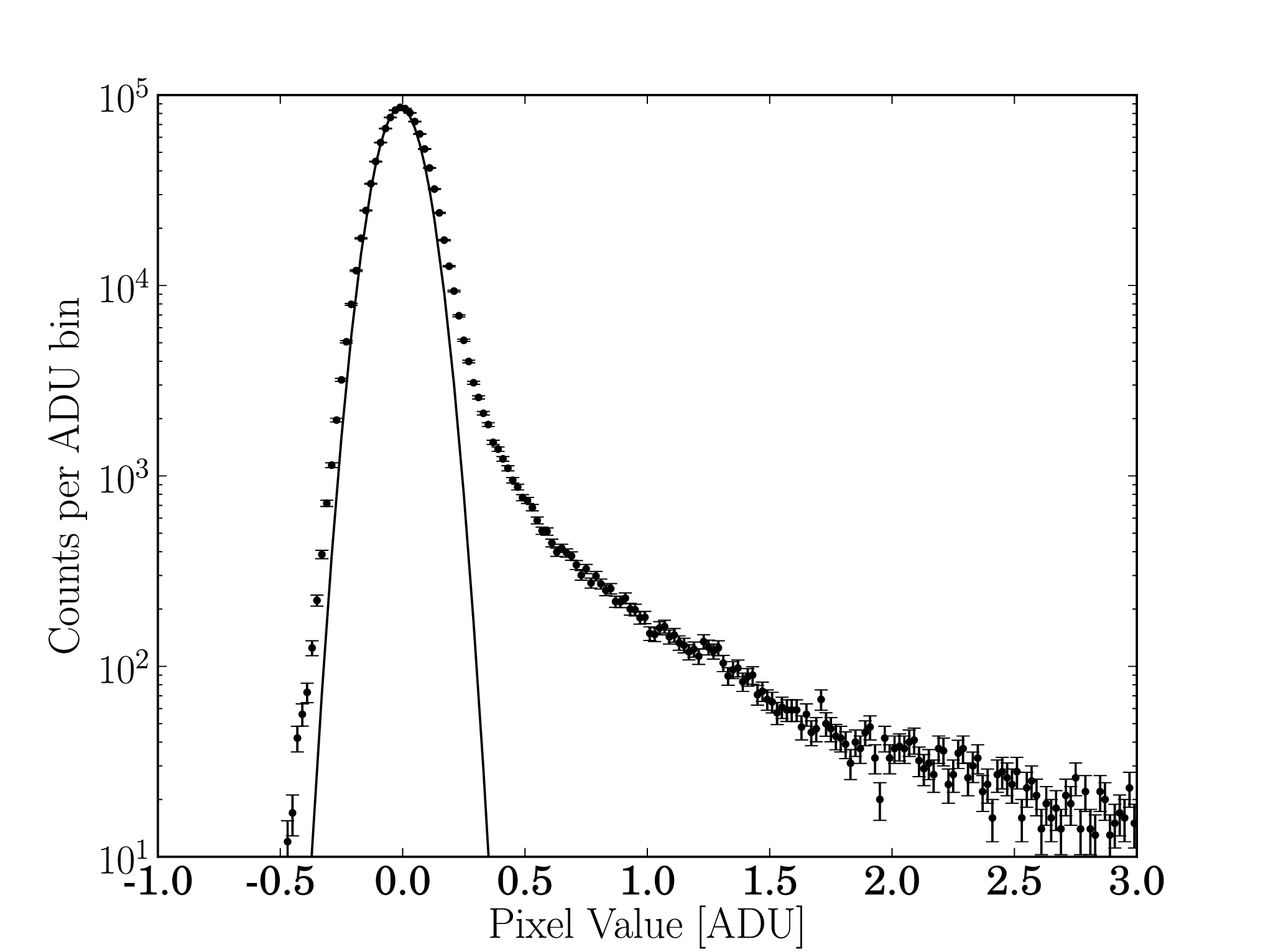} \hspace*{-0.5cm}\includegraphics[width=7.5cm]{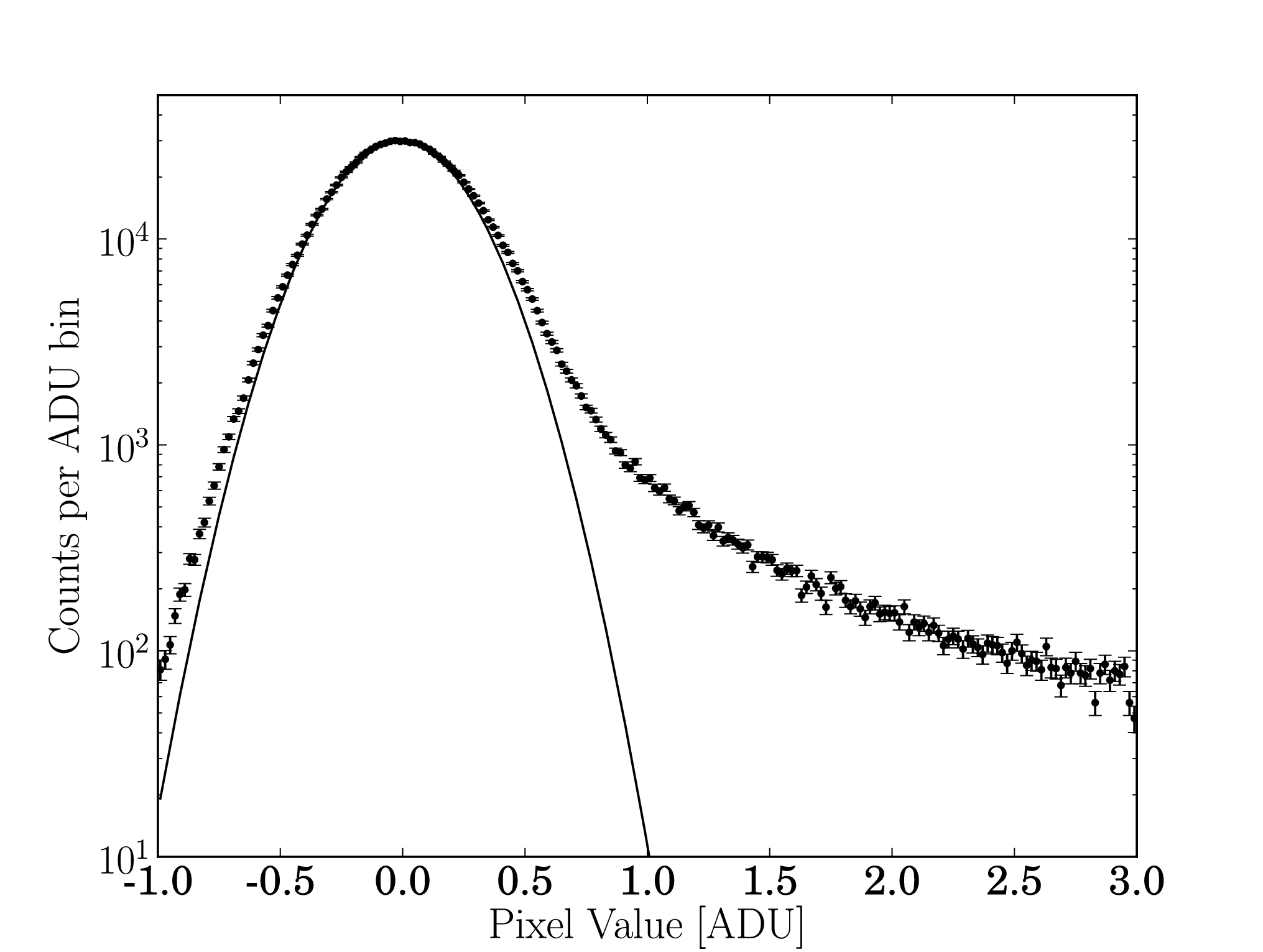}  
  \caption{Logarithmic pixel intensity histograms for the final V-band and I-band CDI images. The Gaussian fits to the background noise correspond to a mean intensity of 0 and a rms noise of about $0.2\%$ of the sky level in both bands. This translates to  $1\sigma$ detections down to  $\sim30$ and $\rm \sim29~mag~arsec^{-2}$ respectively.  The small deviation from perfect parabolae at negative intensities is due to the fact that not all pixels in the final images contain contributions from the same number of CDIs.}
  \label{noise}
\end{figure}

The CDI technique allows us  to obtain  extremely uniform backgrounds over a 4.3 arcmin wide section in RA (1.05 Mpc at the cluster redshift), due to the offset used to nodd the CCD during observations. At scales larger than 4.3 arcmin the background is also very uniform, but information has been lost due to the CDI reduction technique. The logarithm of the frequency distribution of pixels values shown in Figure~\ref{noise} indicates that the background noise is Gaussian with a mean value of 0 ADUs and a dispersion equivalent to a background uniformity of $\sim0.2\%$ of the original sky level in both bands. We remark  that our original  I-band images have a mean sky brightness of $\sim19.5\ mag\ arcsec^{-2}$ while using the CDI technique we were able to detect surface brightness levels down to $I\sim29\rm~ mag~arcsec^{-2}$ at $1\sigma$ above the sky level. Observations were carried out with an instrumental seeing of 0.8 to 1.1 arcsec, and about $40\%$ moon illumination. 

We also have deep R images taken with the FORS2 instrument at VLT in the night of September 22, 2006. They were acquired initially to construct the masks for our spectroscopic observations, but since we had access to the calibration data of that night we also reduced them to have R band magnitudes (in this case we did not use the CDI method).

\begin{table}
\renewcommand{\arraystretch}{0.8}
\caption{CDI Imaging observations for RX J0054.0-2823}                  
\centering                                    
\begin{tabular}{ c c c c}          
\hline                       
	& Night	& Number of & Total Observing \\
	&		& Images      & Time \\     
\hline                                    
 SUSI V & 1 & 14 &  \\
        & 2 & 19 & 9900s \\
 SUSI I & 1 & 24 &  \\
        & 2 & 18 & 12600s \\ 
 FORS2 R & 1 & 6 & 2760s \\
\hline                                   
\end{tabular}
\label{tableob}  
\end{table}

\begin{figure}
  \includegraphics[width=8.6cm]{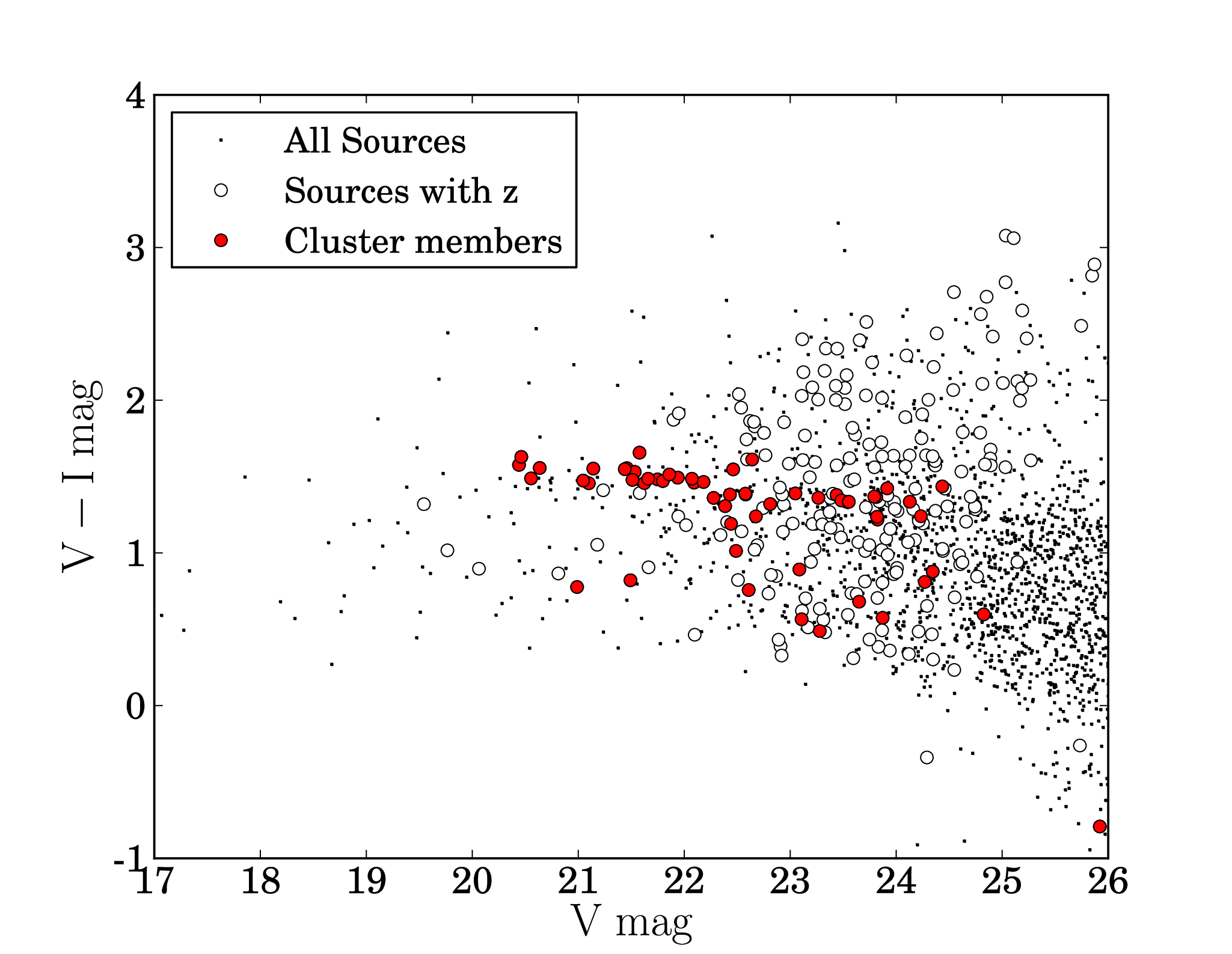}
  \caption{Color-magnitude diagram of the full field of RX J0054.0-2823. Open circles show objects with measured radial velocities, red symbols depict cluster members and points correspond to galaxies with no spectra. The red-sequence of the cluster is clearly appreciated in this figure.}
  \label{cmd}
\end{figure}

We used Sextractor to produce a catalogue with positions and magnitudes of all galaxies in our SUSI V and I images and FORS2 R band. We first run sextractor over the V data. To derive meaningful colours, the I-band and R-band catalogs were created  using the objects and their structural parameters as determined in the V image ({\tt ASSOC} option in Sextractor).  Figure~\ref{cmd} presents the resulting colour-magnitude diagram of the field where different symbols are used to represent objects with measured radial velocities, and members of the central X-ray cluster. The complete catalogue of cluster galaxies with photometry and measured radial velocities in the range of SUSI2 FoV is presented in Table~\ref{cata}.

\subsection{Spectroscopy}

\begin{figure}
  \includegraphics[width=8.6cm]{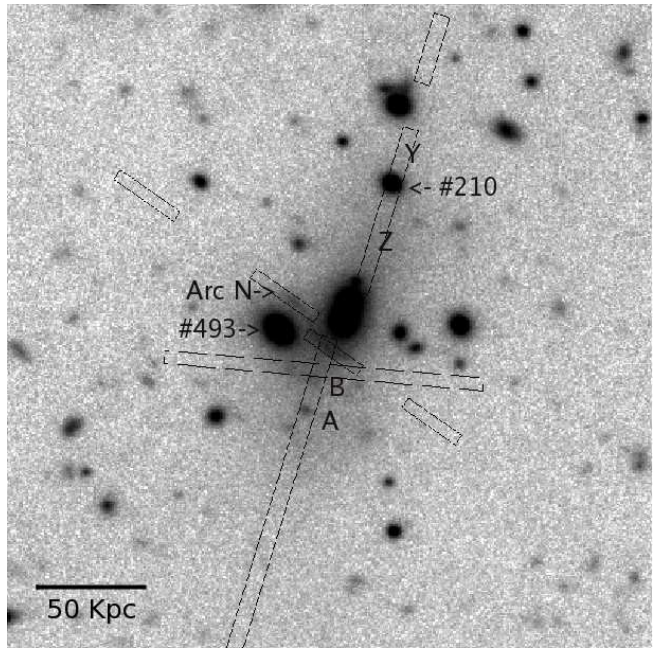}
  \caption{Location of the FORS2 long-slits superposed on the R-band image of the cluster.}
  \label{slits}
\end{figure}

Long-slit and multi-object spectroscopy of the dominant cluster galaxies, the diffuse component, the arc, and about 650 galaxies in the field were obtained using FORS2 on the VLT. Full details about the observations and the data reduction of the individual galaxies are given by \cite[][henceforth Paper~I]{Giraud2010}. Here we present a detailed account of the long-slit data, which is not given in that paper, including particulars relevant to the long-slit observations of the S-shaped arc, the BCG halo and ICL component. Figure~\ref{slits} shows the positions of the slits drawn on our R-band image of the cluster, where the codes that we will use to label the corresponding spectra are identified.

Comparison with our deep images of the BCG halo and ICL shown below (Fig. \ref{ignacio}) show that the outermost slits, which we used to measure the sky, are indeed free of contamination by the BCG halo or the ICL. The two sky positions parallel to the slits on the S-shaped arc were used to correct the spectra for sky features. The figure also shows that the S-shaped arc (that is covered by two slits) suffers from significant contamination by the BCG haloes, which was corrected subtracting the light from the outermost pixels along the corresponding slits of each arc component.

The spectroscopic observations were spread over two periods (P78 and P80) to satisfy the conditions of dark and clear sky with sub-arsecond seeing. In order to match the major and minor axis of the ICL and the symmetric S-shaped arc, rotation angles of $-343^o$, $-85^o$, and $-55^o$ were applied. The slit lengths used for the ICL spectra are  56.5$''$, 32.5$''$, and 24.5$''$. Slit widths are 1.6$''$ for the ICL, and 1$''$ for the arcs. We used the grism 300V with the order sorting filter GG435 for the ICL and the arcs. Thus, the 300V grism covers a wavelength range between $4450 - 8700$\AA\  at a resolution of 112\AA$~{\rm mm}^{-1}$. With the detector used in binned mode the pixel resolution is 3.36~\AA~pixel$^{-1}$. We also used  the 600RI grism for the arc with the same order sorting filter, which covers the range 5120-8450\AA\  at a resolution of 55\AA~${\rm mm}^{-1}$.  The total exposure time for each of the ICL axis is $18 \times 1450s$, and that of the arcs is $12 \times 1450s$ with the 300V grism and $10 \times 1450s$ with the 600RI. The total telescope time including field acquisition, mask positioning, and integration time is 58 hours (1h of telescope time provides $2 \times 1450~s$ of integration). The journal of observations is given in Paper~I. The reader will notice that our slit width of 1.6$''$ is rather small for measurements of very low surface brightness objects. However, given the low richness and low  X-ray luminosity of the cluster, our slit choice was a compromise between sensitivity and spectral resolution.

The data were first reduced by the ESO quality control group who provided us with science products together with calibration data:  master bias (bias and dark levels, read-out noise), master screen flats (high spatial frequency flat, slit function), wavelength calibration spectra from He-Ar lamps, and a set of spectrophotometric standards. We repeated the reduction pipeline starting from bias subtraction, flat fielding, and wavelength calibration. We estimated the sky background on slitlets away from the ICL, either on one side of the ICL slits, or both. Distortion with respect to the columns was measured on the sky line at 5577\AA\ and used to build a 2D sky which was subtracted to the 2D spectrum. The 2D spectra were divided by the CCD response curve derived from spectrophotometric standards, but were not calibrated in flux. Multiple exposures where then aligned and median averaged. The spatial alignment of 2D spectra of the ICL major semi-axis South (ICL-S) was done by using the  spectrum of a faint galaxy, namely object A in Figure~\ref{slits}. For the semi-major axis of the  northern part of the ICL (ICL-N marked Y in Figure~\ref{slits}) we used the profile of galaxy \#210. Galaxy A is a distant background object, and galaxy 210 is a cluster galaxy possibly in the ICL. For the ICL minor axis we used the spatial profiles of the ICL spectra obtained by coadding columns from 5800\AA\ to  7550\AA\ . The same procedure was done for both components (slits) of the S-shaped arc.   

\begin{table*}
\renewcommand{\arraystretch}{0.7}
\renewcommand{\baselinestretch}{0.6}
\renewcommand{\tabcolsep}{1.5mm}
              
\centering
\begin{minipage}{160mm}
\caption{Catalogue of redshift and photometric properties of cluster galaxies in the field covered by SUSI2. The data for all the 95 cluster galaxies will be given in an electronic catalogue.}
\begin{tabular}{ l l l l l l l l l l l l l}          
\hline\small                              
& Number  	& RA 		& DEC    & $r$\footnote{Distance to the center of the cluster} 	     	& $z$ 		& $m_\mathrm{R}$\footnote{Because of the crowding in the central area of the images, some galaxies are left without photometric data}  	& $V_\mathrm{rad}$ 		& $\sigma(V_\mathrm{rad})$ 	& $m_\mathrm{V}$ 	& $\sigma(m_\mathrm{V})$ 	& $V-I$ 	& $\sigma(V-I)$ \\
& (Paper I) 	& (deg)		& (deg)  & (arcsec)  	& 	    	& (mag)			& ($\mathrm{km~s^{-1}}$)	& ($\mathrm{km~s^{-1}}$)	& (mag)			& (mag)				& (mag)		& (mag) \\ \hline
\small 1 & 9 &   13.502186 & -28.406406 &   46.0 &  0.29266 & 21.60 &   75362 &  42 & 22.76 &  0.03 &  1.29 &  0.03 \\ 
2 & 12 &  13.495841 & -28.391250 &   81.5 &  0.29730 & 21.97 &   76369 &  22 & 22.64 &  0.03 &  0.75 &  0.04 \\ 
3 & 30 &  13.580301 & -28.435437 &  262.2 &  0.29032 & 21.08 &   74853 &  49 & 22.01 &  0.02 &  1.01 &  0.03 \\ 
4 & 46 &  13.533176 & -28.388791 &   85.7 &  0.29293 & 20.25 &   75421 &  20 & 20.91 &  0.01 &  0.89 &  0.02 \\ 
5 & 50 &  13.503106 & -28.404664 &   41.8 &  0.28861 & 22.15 &   74480 &  42 & 23.38 &  0.04 &  1.20 &  0.05 \\ 
6 & 71 &  13.530753 & -28.417693 &   76.5 &  0.29381 & 22.16 &   75612 &  40 & 23.67 &  0.05 &  1.19 &  0.06 \\ 
7 & 132 &  13.444300 & -28.375612 &  273.0 &  0.29340 & 20.17 &   75523 &  37 & 21.08 &  0.01 &  1.45 &  0.02 \\ 
8 & 139 &  13.497322 & -28.389076 &   81.9 &  0.29237 & 20.53 &   75299 &  32 & 21.64 &  0.02 &  1.47 &  0.02 \\ 
9 & 140 &  13.519294 & -28.400171 &   21.1 &  0.28781 & 21.96 &   74305 &  20 & 22.70 &  0.02 &  0.93 &  0.03 \\ 
10 & 141 &  13.508329 & -28.384237 &   74.0 &  0.29477 & 21.00 &   75821 &  34 & 22.06 &  0.02 &  1.37 &  0.02 \\ 
11 & 142 &  13.503175 & -28.370500 &  126.8 &  0.29481 &  0.00 &   75829 &  22 & 24.07 &  0.05 &  1.36 &  0.06 \\ 
12 & 147 &  13.560627 & -28.416407 &  171.5 &  0.29223 & 18.56 &   75269 &  33 & 19.44 &  0.01 &  1.12 &  0.01 \\ 
13 & 153 &  13.539844 & -28.375597 &  136.0 &  0.29455 & 23.09 &   75773 &  40 & 24.13 &  0.07 &  1.02 &  0.08 \\ 
14 & 170 &  13.472959 & -28.422650 &  164.9 &  0.28994 & 22.76 &   74770 &  25 & 23.98 &  0.06 &  1.27 &  0.07 \\ 
15 & 177 &  13.513872 & -28.391570 &   44.1 &  0.28884 & 20.94 &   74530 &  31 & 22.01 &  0.02 &  1.34 &  0.02 \\ 
16 & 180 &  13.511961 & -28.369568 &  123.6 &  0.28866 & 21.90 &   74491 &  36 & 22.93 &  0.03 &  1.34 &  0.03 \\ 
17 & 181 &  13.524153 & -28.366597 &  138.1 &  0.29389 & 20.62 &   75630 &  28 & 21.53 &  0.01 &  1.38 &  0.02 \\ 
18 & 189 &  13.475854 & -28.398279 &  141.2 &  0.29694 & 20.78 &   76291 &  43 & 21.92 &  0.03 &  1.34 &  0.03 \\ 
19 & 200 &  13.484198 & -28.373495 &  154.7 &  0.29313 & 21.52 &   75465 &  20 & 22.10 &  0.02 &  0.81 &  0.03 \\ 
20 & 201 &  13.491214 & -28.373469 &  138.0 &  0.29124 & 20.85 &   75053 &  33 & 23.20 &  0.04 &  1.32 &  0.05 \\ 
21 & 209 &  13.509186 & -28.395949 &   34.5 &  0.28559 & 22.33 &   73819 &  20 & 22.85 &  0.02 &  0.57 &  0.04 \\ 
22 & 210 &  13.513265 & -28.400198 &   13.9 &  0.29316 & 19.30 &   75471 &  38 & 20.43 &  0.01 &  1.49 &  0.01 \\ 
23 & 211 &  13.512869 & -28.404425 &    7.0 &  0.28644 & 20.34 &   74005 &  50 & 21.24 &  0.01 &  1.35 &  0.01 \\ 
24 & 212 &  13.520299 & -28.415043 &   45.3 &  0.29413 & 21.44 &   75682 &  20 & 22.60 &  0.02 &  1.39 &  0.03 \\ 
25 & 213 &  13.519221 & -28.420006 &   60.6 &  0.29259 & 21.98 &   75347 &  48 & 23.36 &  0.03 &  1.29 &  0.04 \\ 
26 & 214 &  13.548498 & -28.422960 &  139.9 &  0.29266 & 19.80 &   75362 &  20 & 20.52 &  0.01 &  0.85 &  0.01 \\ 
27 & 216 &  13.506242 & -28.438556 &  128.8 &  0.28859 &  0.00 &   74475 &  20 & 24.27 &  0.05 &  0.46 &  0.08 \\ 
28 & 224 &  13.559791 & -28.451988 &  237.6 &  0.29004 & 22.05 &   74792 &  20 & 24.02 &  0.15 &  1.10 &  0.17 \\ 
29 & 231 &  13.509300 & -28.376835 &   98.9 &  0.29279 & 19.65 &   75391 &  54 & 20.64 &  0.01 &  1.43 &  0.01 \\ 
30 & 235 &  13.499133 & -28.436700 &  131.1 &  0.29384 & 21.00 &   75619 &  50 & 22.18 &  0.02 &  1.26 &  0.03 \\ 
31 & 238 &  13.509125 & -28.445594 &  151.9 &  0.29179 & 21.65 &   75173 &  65 & 23.31 &  0.04 &  1.24 &  0.04 \\ 
32 & 252 &  13.505991 & -28.396574 &   40.7 &  0.29656 & 21.81 &   76209 &  55 & 22.83 &  0.03 &  1.22 &  0.03 \\ 
33 & 281 &  13.548612 & -28.400804 &  122.6 &  0.28908 & 22.93 &   74582 &  27 & 23.88 &  0.06 &  0.73 &  0.08 \\ 
34 & 307 &  13.542390 & -28.436479 &  154.3 &  0.29220 & 22.34 &   75262 &  29 & 23.30 &  0.04 &  0.77 &  0.06 \\ 
35 & 321 &  13.472180 & -28.382140 &  171.7 &  0.29288 & 22.34 &   75410 &  41 & 23.52 &  0.05 &  1.38 &  0.06 \\ 
36 & 322 &  13.494444 & -28.387932 &   92.6 &  0.29350 & 22.73 &   75545 &  25 & 23.51 &  0.04 &  0.73 &  0.06 \\ 
37 & 441 &  13.582495 & -28.451167 &  297.8 &  0.29001 & 20.27 &   74785 &  71 & 21.81 &  0.05 &  1.31 &  0.06 \\ 
38 & 446 &  13.433374 & -28.389027 &  297.5 &  0.29278 & 20.87 &   75389 &  83 & 22.01 &  0.02 &  1.52 &  0.03 \\ 
39 & 447 &  13.430624 & -28.392443 &  305.4 &  0.29278 & 20.95 &   75389 &  83 & 22.00 &  0.02 &  1.29 &  0.03 \\ 
40 & 486 &  13.520790 & -28.368487 &  128.9 &  0.29533 & 20.13 &   75942 &  73 & 21.21 &  0.01 &  1.50 &  0.01 \\ 
41 & 490 &  13.522448 & -28.394964 &   42.3 &  0.29030 & 19.93 &   74849 &  89 & 21.10 &  0.01 &  1.47 &  0.01 \\ 
42 & 491 &  13.517445 & -28.394719 &   34.1 &  0.29240 & 19.54 &   75306 &  77 & 20.69 &  0.01 &  1.49 &  0.01 \\ 
43 & 492 &  13.502883 & -28.387377 &   72.8 &  0.29203 & 20.29 &   75225 &  78 & 21.36 &  0.01 &  1.43 &  0.02 \\ 
58 & 493 &  13.516765 & -28.404248 &   13.1 &  0.29350 & 18.26 &   75545 &  42 & 19.27 &  0.01 &   -   &   -   \\
44 & 494 &  13.511110 & -28.404158 &   13.0 &  0.29436 & 19.37 &   75732 & 101 & 20.72 &  0.01 &  1.45 &  0.01 \\ 
45 & 496 &  13.494031 & -28.399643 &   75.9 &  0.29068 & 20.03 &   74931 & 108 & 21.15 &  0.01 &  1.46 &  0.02 \\ 
46 & 497 &  13.479428 & -28.393846 &  131.9 &  0.29458 & 19.45 &   75780 &  63 & 20.55 &  0.01 &  1.43 &  0.01 \\ 
47 & 498 &  13.495845 & -28.427411 &  108.8 &  0.29431 & 19.88 &   75721 &  75 & 21.04 &  0.01 &  1.47 &  0.01 \\ 
48 & 500 &  13.451707 & -28.409518 &  227.7 &  0.29382 & 19.83 &   75615 &  90 & 20.96 &  0.01 &  1.55 &  0.01 \\ 
49 & 501 &  13.494418 & -28.450216 &  182.4 &  0.29272 & 20.13 &   75376 & 119 & 21.37 &  0.04 &  1.41 &  0.04 \\ 
50 & 506 &  13.525435 & -28.412273 &   49.3 &  0.29297 & 20.74 &   75430 &  49 & 21.80 &  0.02 &  1.36 &  0.02 \\
59 & 509 &  13.515700 & -28.403790 &    0.0 &  0.29290 & 17.33 &   75415 &  43 & 18.29 &  0.01 &   -   &   -   \\
69 & 510 &  13.51458  & -28.403490 &    0.7 &  0.29340 &   -   &   75523 &  72 &   -   &   -   &   -   &   -   \\
51 & 512 &  13.513087 & -28.397975 &   21.7 &  0.29256 & 19.14 &   75341 &  76 & 20.26 &  0.01 &  1.45 &  0.01 \\ 
52 & 513 &  13.501408 & -28.393905 &   59.6 &  0.28637 & 20.16 &   73990 &  67 & 21.27 &  0.01 &  1.43 &  0.02 \\ 
53 & 538 &  13.523675 & -28.389112 &   61.9 &  0.29084 & 21.82 &   74966 &  36 & 22.98 &  0.03 &  1.29 &  0.03 \\ 
54 & 548 &  13.457036 & -28.425231 &  221.5 &  0.28902 & 23.68 &   74569 &  38 & 24.67 &  0.08 &  0.75 &  0.11 \\ 
55 & 1009 &  13.523528 & -28.426816 &   88.8 &  0.29040 & 18.60 &   74870 &   724 & 19.76 &  0.01 &  1.51 &  0.01 \\ 
56 & 1018 &  13.491513 & -28.389769 &   97.5 &  0.29250 & 19.62 &   75328 &   723 & 20.79 &  0.01 &  1.46 &  0.02 \\ 
57 & 1020 &  13.481431 & -28.403676 &  119.7 &  0.29390 & 20.71 &   75632 &   726 & 21.81 &  0.02 &  1.30 &  0.02 \\ 
\hline                                        
\end{tabular}
\end{minipage}
\label{cata} 
\end{table*}
\normalfont
\normalfont

\section{Results}
\label{results}

\subsection{Global properties of the cluster}
\label{SECcluster}
 
\begin{figure}
  \includegraphics[width=8.6cm]{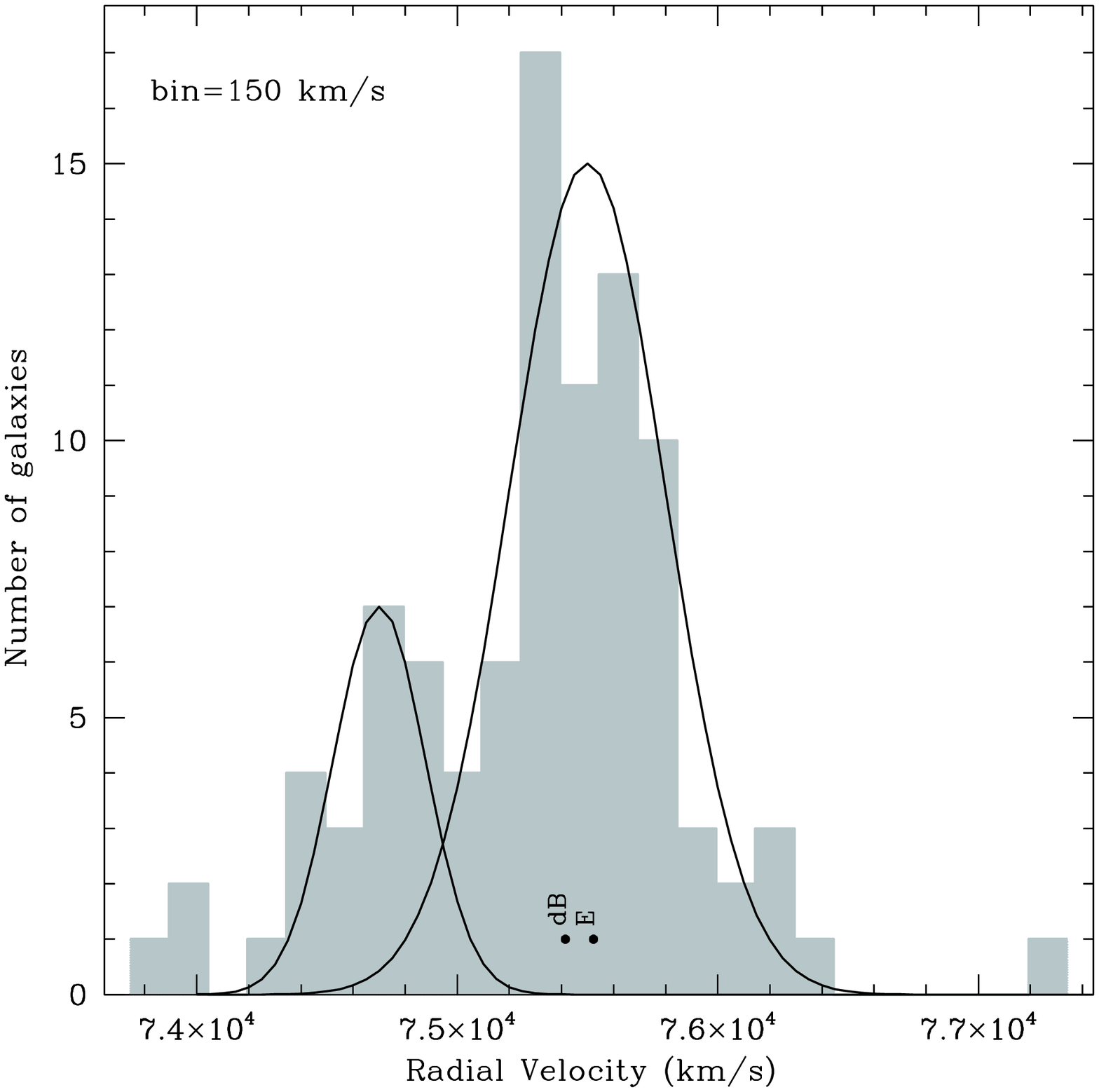}
  \caption{Radial velocity histogram for 95 probable cluster members with a bin size of 150 km/s. The double-peaked structure of the histogram does not depend on bin size.}
  \label{radialv}
\end{figure}

\begin{figure}
  \includegraphics[width=8.6cm]{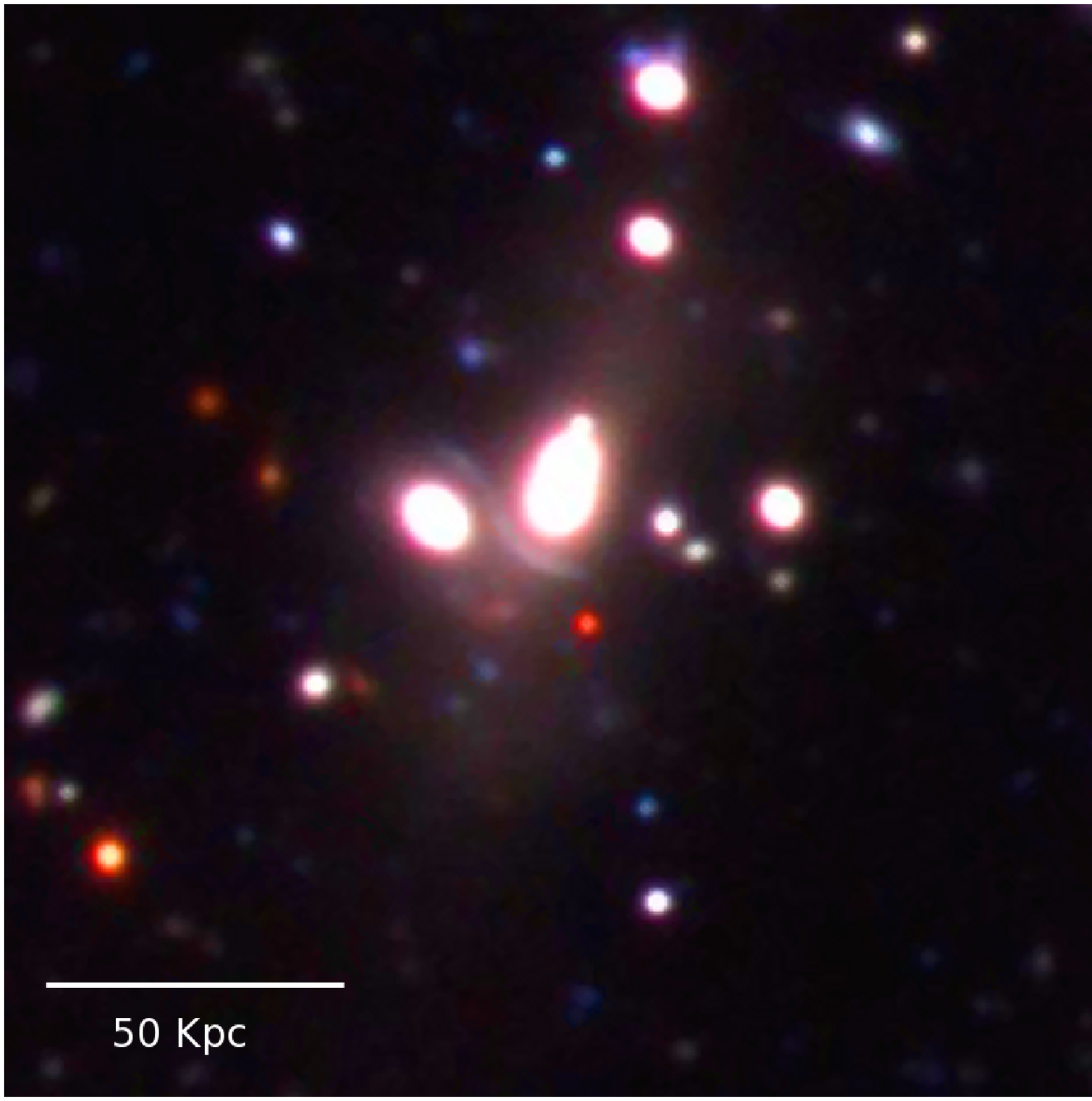}
  \caption{VRI colour composite of the central part of the cluster showing very clearly the S-shaped arc, the dumbbell (BCG) and the third close brightest elliptical galaxy (\#493) to the East.}
  \label{cen}
\end{figure}

Out of the $\sim650$ galaxy spectra that we collected in the field of the cluster, 95 had redshifts in the range $0.283<z<0.303$ that we considered to be possible members of the cluster. Figure~\ref{radialv} shows the radial velocity histogram of the cluster for a bin size of $150~km~s^{-1}$ that shows two distinct peaks. The strongest peak, which contains close to 90\% of the galaxies, is centered very precisely on the average radial velocity of the three brightest cluster members. The second peak corresponds to galaxies with negative velocities relative to the brightest members. This could be interpreted as evidence for the infall of a group of galaxies along a broad filament just behind the cluster onto the cluster centre defined by the brightest galaxies.  However, the spatial distribution of the putative infalling galaxies is not distinct from that of the main-peak galaxies, although none of the brightest galaxies in the cluster corresponds to that group. Figure~\ref{cen} shows a composite colour image of the central part of the cluster where the three dominant central elliptical galaxies are clearly seen together with the S-shaped arc. Two of the central ellipticals form a very close dumbbell system that, however, does not show the typical tidal distortions in their envelopes that characterize well known dumbbells in groups or nearby clusters (such as NGC 4782/4783 \citep{Borne1993} or A3391 \citep{Govoni2000}). The higher redshift of our cluster (compared to previously studied systems) and photometric effects from the third elliptical galaxy and the S-shaped arc, however, may make it hard to distinguish faint distortions. 

\begin{figure}
  \includegraphics[width=8.6cm]{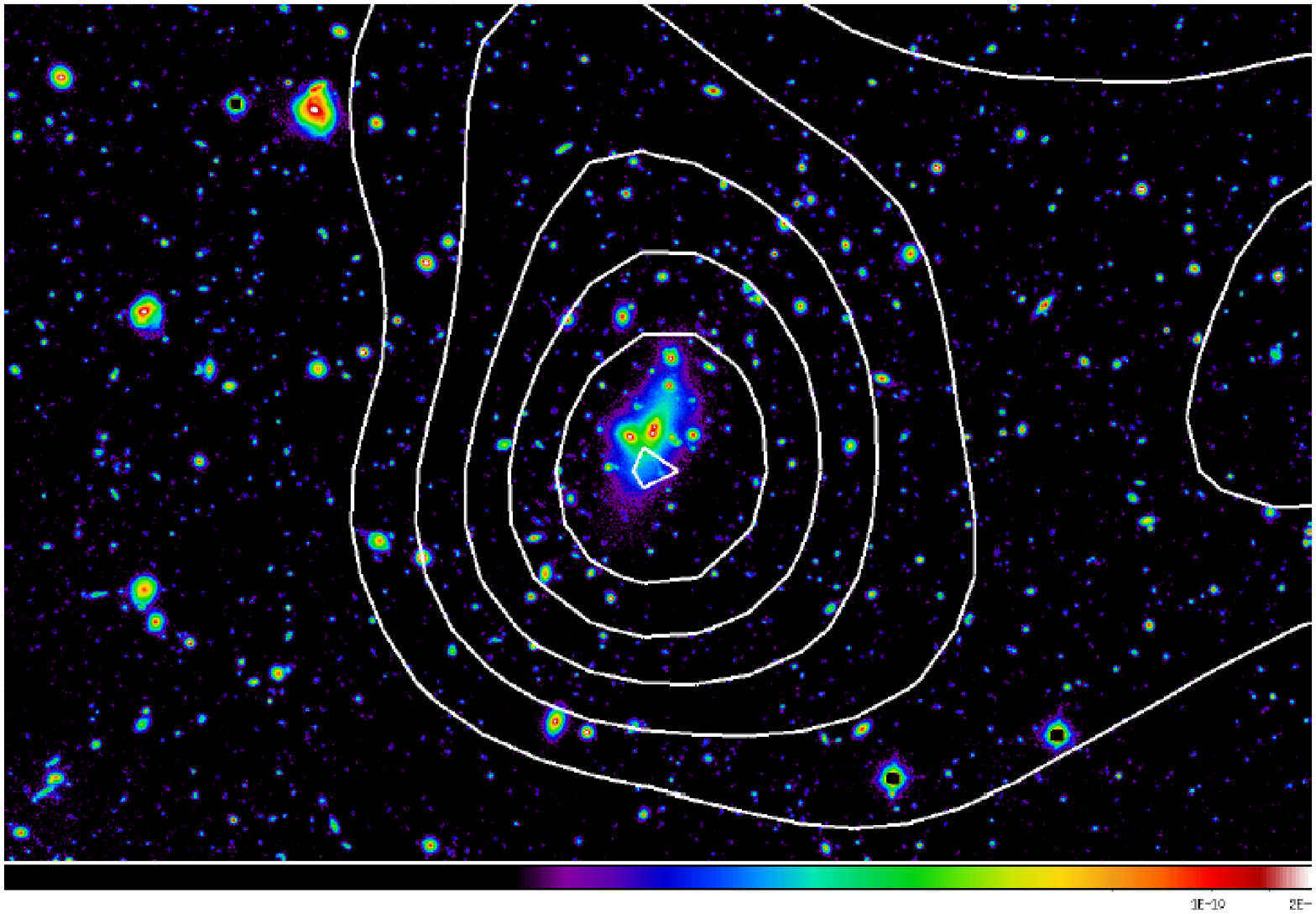}
  \caption{ROSAT X-ray contours superposed on a false colour image of the cluster. The X-ray contours are fairly symmetric with some elongation roughly in the direction of the ICL in the outer parts.}
  \label{xrays}
\end{figure}

We cannot exclude that one of the components of the dumbbell system originally had a radial velocity consistent with the {\it infalling}  peak of the histogram, and that the passage close to the other component and the cluster centre could have deflected its original motion thus masking its dynamical origin, but the evidence for such a major merger event is rather weak: the X-ray contours shown in Figure~\ref{xrays}  have a slight elongation along the dumbbell axis although ROSAT does not detect any substructure. This may be due to the resolution and sensitivity of the ROSAT data  for this relatively faint cluster in X-rays, although the presence of a relaxed, smooth cluster structute does not necessarily imply that the cluster is not in an advanced stage of merging  \citep{Brough2008}.  The third E-galaxy (\#493) also has a small velocity difference relative to the db, but not knowing its real position along the line of sight, it remains rather difficult to ascertain its dynamical role in the cluster centre.

The possibility that the double peaked radial velocity histogram reflects a merger of two clusters was further investigated by Monte-Carlo simulations of a single peaked Gaussian trying to detect secondary  peaks produced by statistical fluctuations. We generated 100,000 galaxies with the overall (single) Gaussian velocity dispersion of the cluster, drawing 1000 random selections of 100 galaxies in each experiment. We then searched for resulting bimodal distributions applying the following conditions to define a statistically significant secondary peak:

\begin{enumerate}
\item Reduced $\chi^{2}$ difference $\geq 1$ ($\chi^2_{single}-\chi^2_{bimod} \geq 1 $)
\item Peak velocity differences $\geq 400$ km/s ($\| \mu_{single} -\mu_{bimod} \| \geq 400 $)
\item Individual amplitudes $\geq 2.5 $ (At least 3 objects in a Gaussian)
\item Minimum FWHM for each Gaussian $\geq 200 km s^{-1}$
\item Amplitudes ratio $\geq 0.25$ ($A_{bimod1(2)}/A_{bimod2(1)} \geq 0.25$)
\end{enumerate}

Using bin sizes in the range $180-250\rm~ km~s^{-1}$ we found that the probabilities of finding spurious bi-modality varied between 16\% and 19\%.  Therefore, there is an 80\% probability that our cluster underwent a significant merger in the past, the few faint galaxies that we presently see infalling onto the cluster centre are the last remnants of  the infalling smaller cluster. Thus, Table~\ref{props} presents the global properties of the cluster obtained using all the gallaxies in our catalog (One Gaussian), or fitting two Gaussians to the radial velocity histogram and retaining the parameters of the main peak to compute the properties of the cluster.

\begin{table}
\caption{Global properties of RX J0054.0-2823}                   
\centering                                   
\begin{tabular}{ l l}          
\hline\hline                        
Parameter & Value \\     
\hline                                    
   X-ray luminosity		& 	$L_X=0.200-0.398\times10^{14}~\mathrm{ergs~s^{-1}}$		\\
     Virial Radius			&	$R_{vir}=0.964~\mathrm{Mpc}$							\\
   Virial  Mass 			&	$ M_{vir}=7.1\times10^{13}~M_{\sun}$				\\
    Total Luminosity 		& 	$L_{vir,V}=8.08\times10^{11}~L_{\sun,V}$				\\
   Mass-to-light ratio		& 	$(M/L)_V=88~(M_{\sun}/L_{\sun})_{V}$	\\
   {\bf One Gaussian}		& 											\\
   Velocity dispersion		&	$\sigma=496\pm60~\mathrm{km~s^{-1}}$						\\
   $R_{500}$          		&	660 Kpc 									\\
   BCG+ICL fraction		&	$0.57\pm0.07$							\\
   {\bf Two Gaussians}	&										\\
   Velocity dispersion		&	$\sigma=328~\mathrm{km~s^{-1}}$						\\
   BCG+ICL fraction		&	$0.41\pm0.06$								\\
   $R_{500}$			& 	413 Kpc									\\
\hline
\end{tabular}
\label{props}
\end{table}

\begin{figure}
  \includegraphics[width=8.6cm]{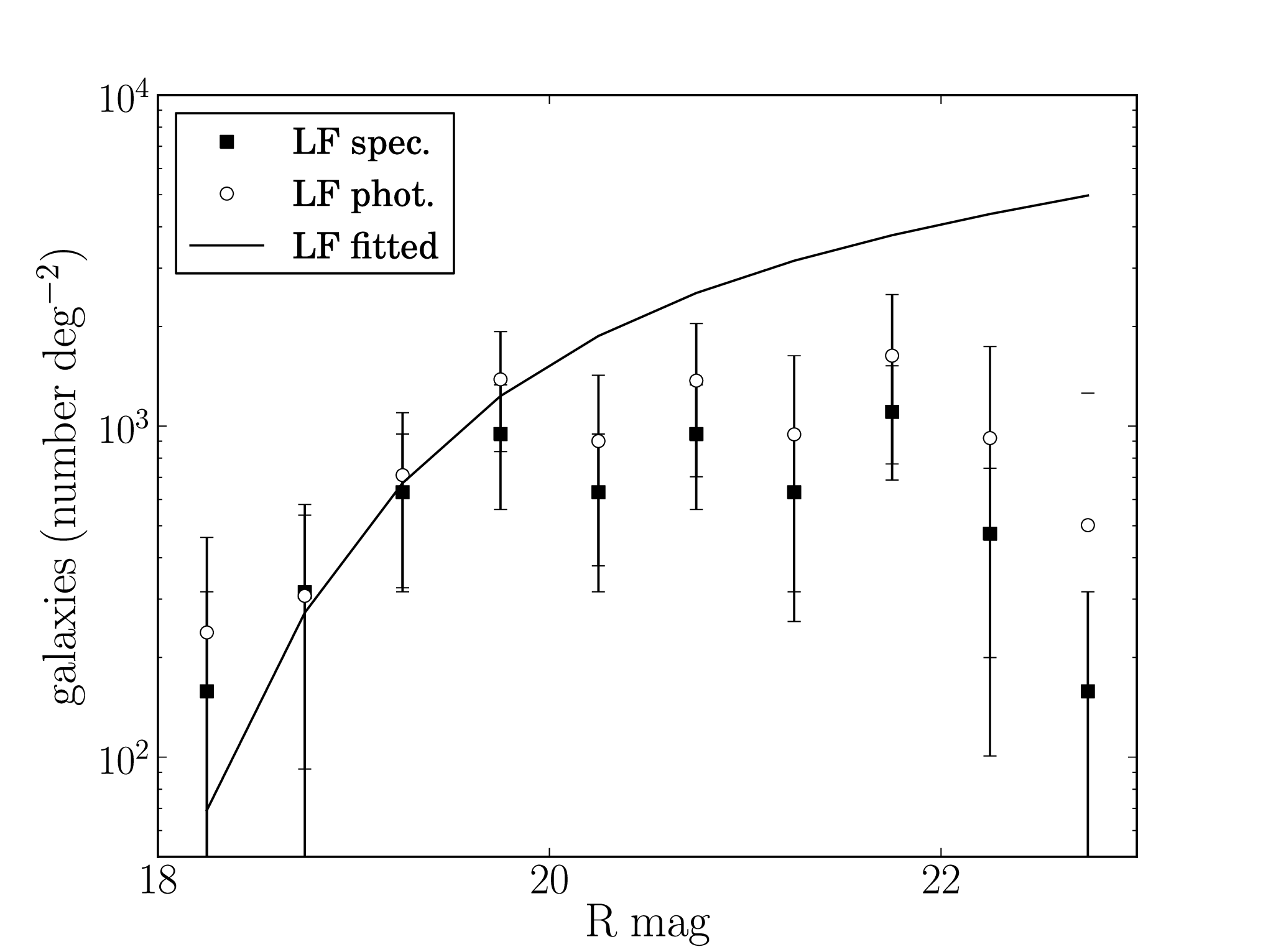}
  \caption{Luminosity function of galaxies in the cluster determined using two sets of data: spectroscopically confirmed cluster members (black dots;  spec), and statistical counts using photometry on a WFI R-band image as described in the text  (green dots; phot).  The solid line shows the fit of a standard cluster luminosity-function to the 4 brightest bins of the spectroscopic data.}
  \label{lf}
\end{figure}

We computed the total luminosity of the cluster summing the light of all cluster members within a well defined radius.  However, this leads us to underestimate the luminosity because our spectroscopic catalogue is not complete.  We thus used an R-band image taken with WFI on the La Silla ESO/MPIA 2.2m telescope to determine the luminosity function of the cluster corrected for field contamination. The result is shown in Figure~\ref{lf} where we compare the luminosity function of the galaxies with spectroscopy to that from the WFI R-band photometry. The solid line shows the fit of a standard 2-parameter cluster luminosity function ($\alpha=-1.21;~M_{*,R}=-21.14$, see \citet{Christlein2003}) adjusted to fit the  four brightest (R=18 to R=19.5) bins from the spectroscopic data.  Thus, we calculate the total luminosity of the cluster by integrating the fitted luminosity function out to faint magnitudes. This is done independently for each value of $r_{500}$ listed in Table~\ref{props}, and we obtain the V and I magnitudes assuming (V-R)=1.0 and (V-I)=1.6.

\subsection{Photometric properties of the ICL}
\label{photr500}

\begin{figure}
  \includegraphics[width=8.6cm]{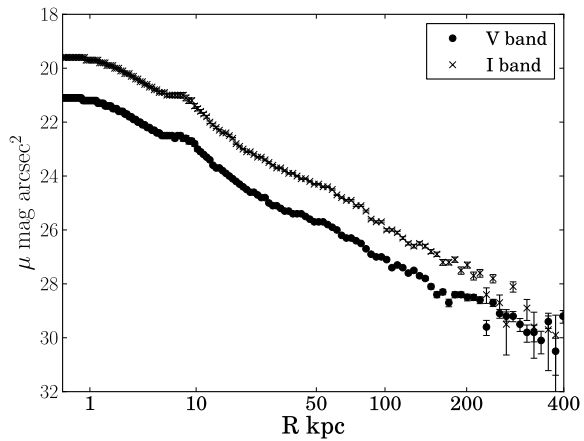}
  \caption{V and I surface brightness profiles in an $r^{-1/4}$ scale. Error bars indicate the excellent quality of the background subtraction, which have errors of equivalent to 0.2\% of the sky. Even at very low levels the errors are dominated by contamination by objects in the field.}
  \label{profis}
\end{figure}

Because of cosmological dimming, the detection of the ICL in clusters should be much easier at low redshifts than at the redshift of our cluster. However, we find that the disadvantage is largely compensated by the fact that intermediate redshift clusters are more compact allowing extremely accurate sky subtraction and also minimizing contamination by Galactic Cirrus that plague investigations of the nearest clusters \citep[e.g.]{Rudick2009}. In the surface brightness profiles presented in Figure~\ref{profis} we can detect the ICL down to at least $\mu_\mathrm{V}=30~\mathrm{mag~arcsec^{-2}}$ and radial distances of more than 350~kpc, showing that can properly characterize the ICL component for this cluster at $z\sim0.3$.  

In order to characterize the properties of the ICL, and in particular the fraction of the total stellar light of the cluster that is contributed by the ICL, we need to define a radius within which we will integrate these properties. To facilitate comparison with previous investigations, and in particular with the work of \citet[][henceforth GZZ]{Gonzalez2007}, we will use $R_{500}$, the radius at which the cluster density is 500 times the critical density of the Universe ($\Omega_c=5\times10^{-30} g~cm^{-3}$). GZZ used the sample of clusters from \citet{Vikhlinin2006} to calibrate the relation between the velocity dispersion of the clusters ($\sigma$) and $R_{500}$ and thus to infer the radius from the velocity dispersion.  Again for consistency, we will use the same calibration to determine the relevant parameters of our cluster, with the caveat that we have two possible values of $\sigma$ as indicated in Table~\ref{props}. We thus infer two different values for $R_{500}$, which we will preserve in all computations.

\begin{figure*}
  \resizebox{\hsize}{!}{\includegraphics{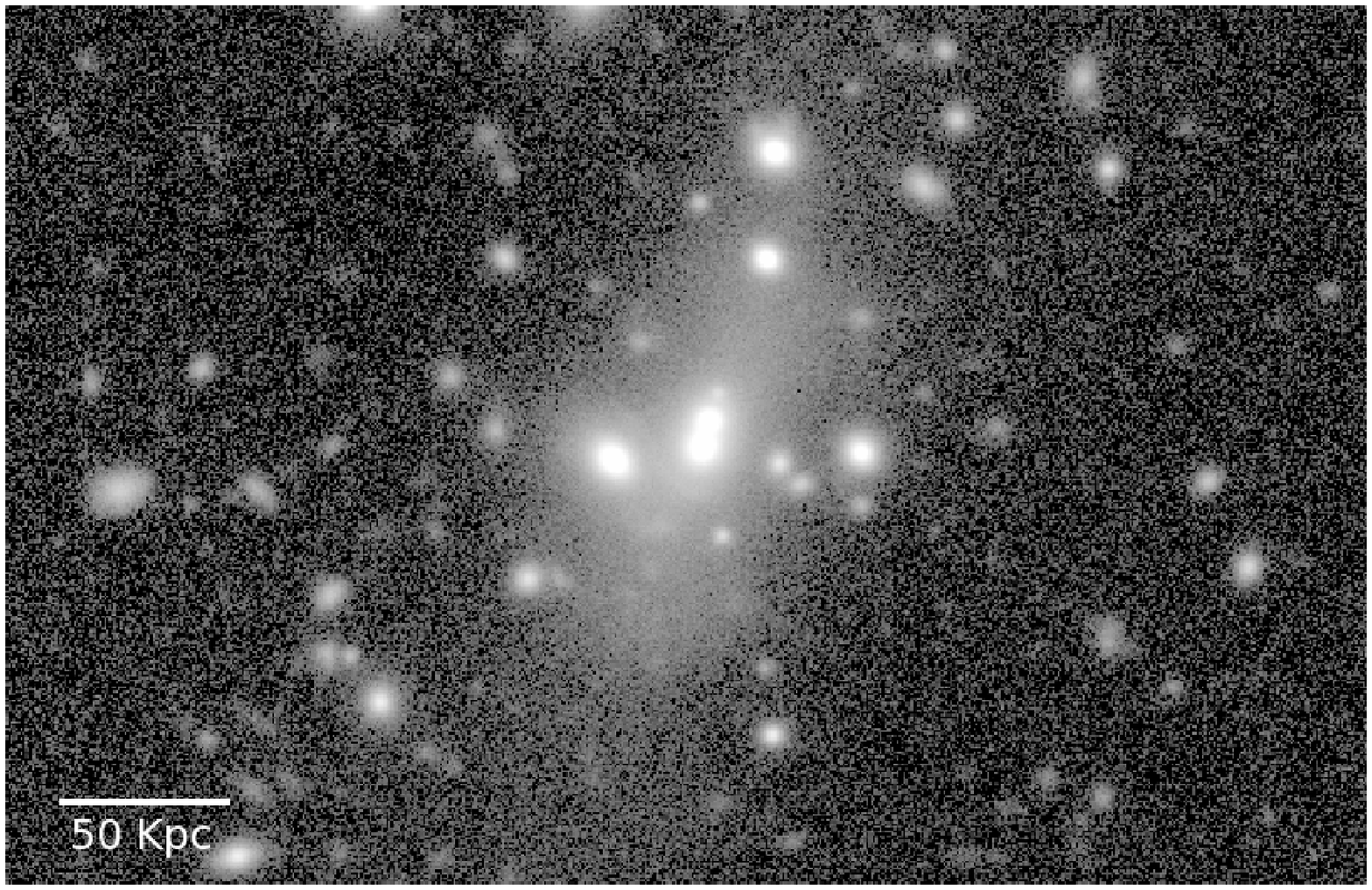}\hspace*{0.1cm}\includegraphics{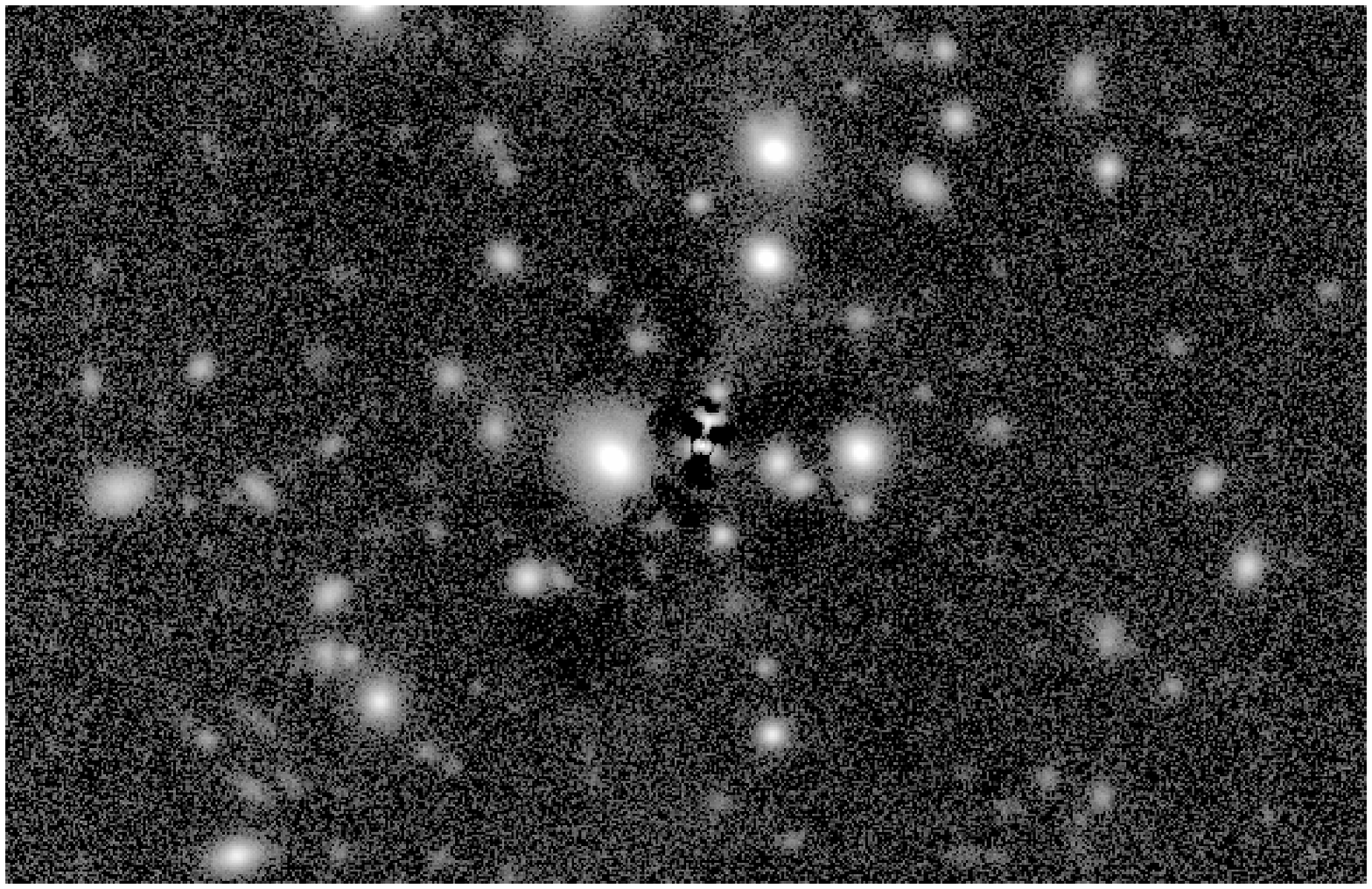}}
  \vspace*{0.1cm}
  \resizebox{\hsize}{!}{\includegraphics{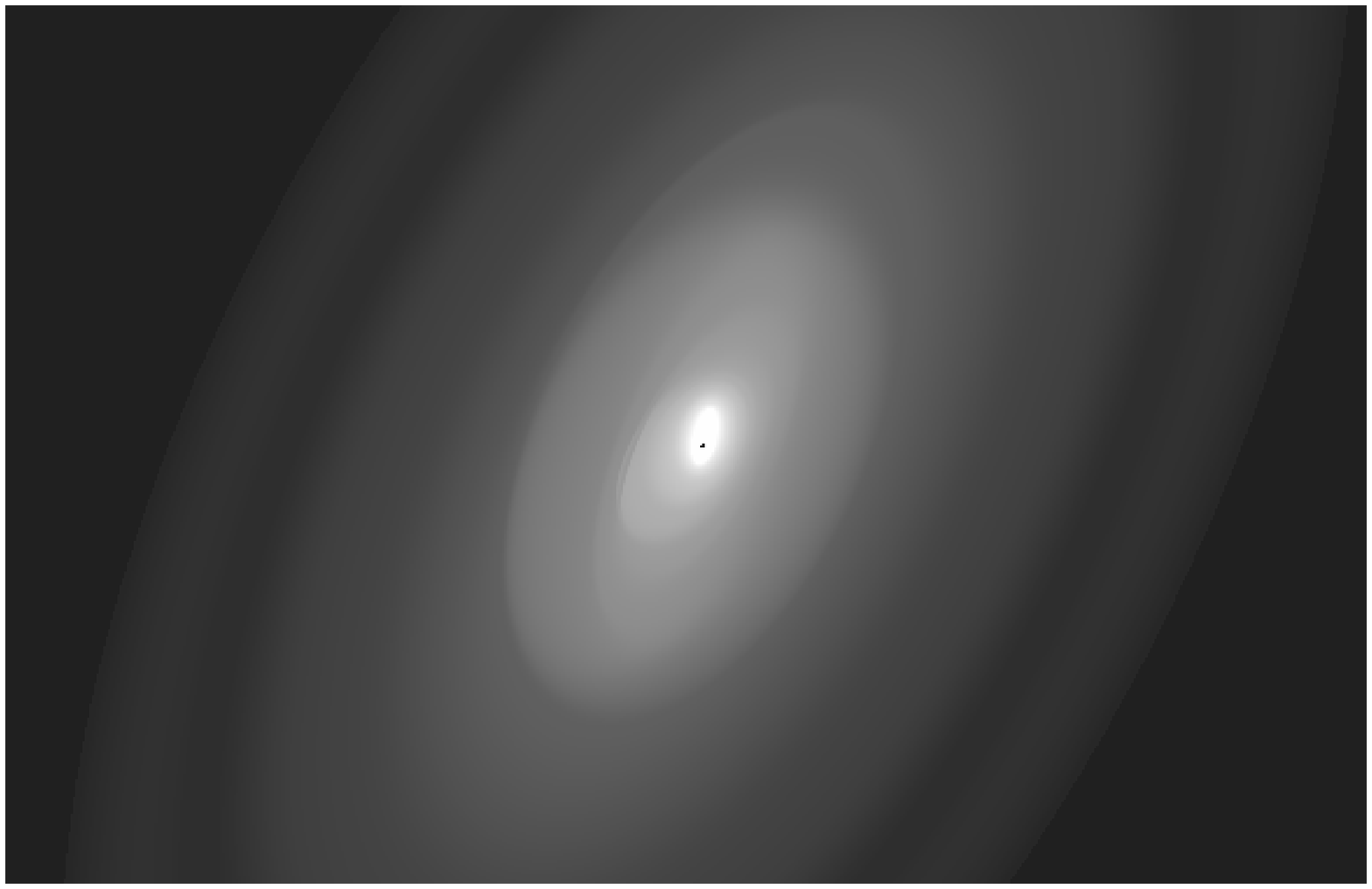}\hspace*{0.1cm}\includegraphics{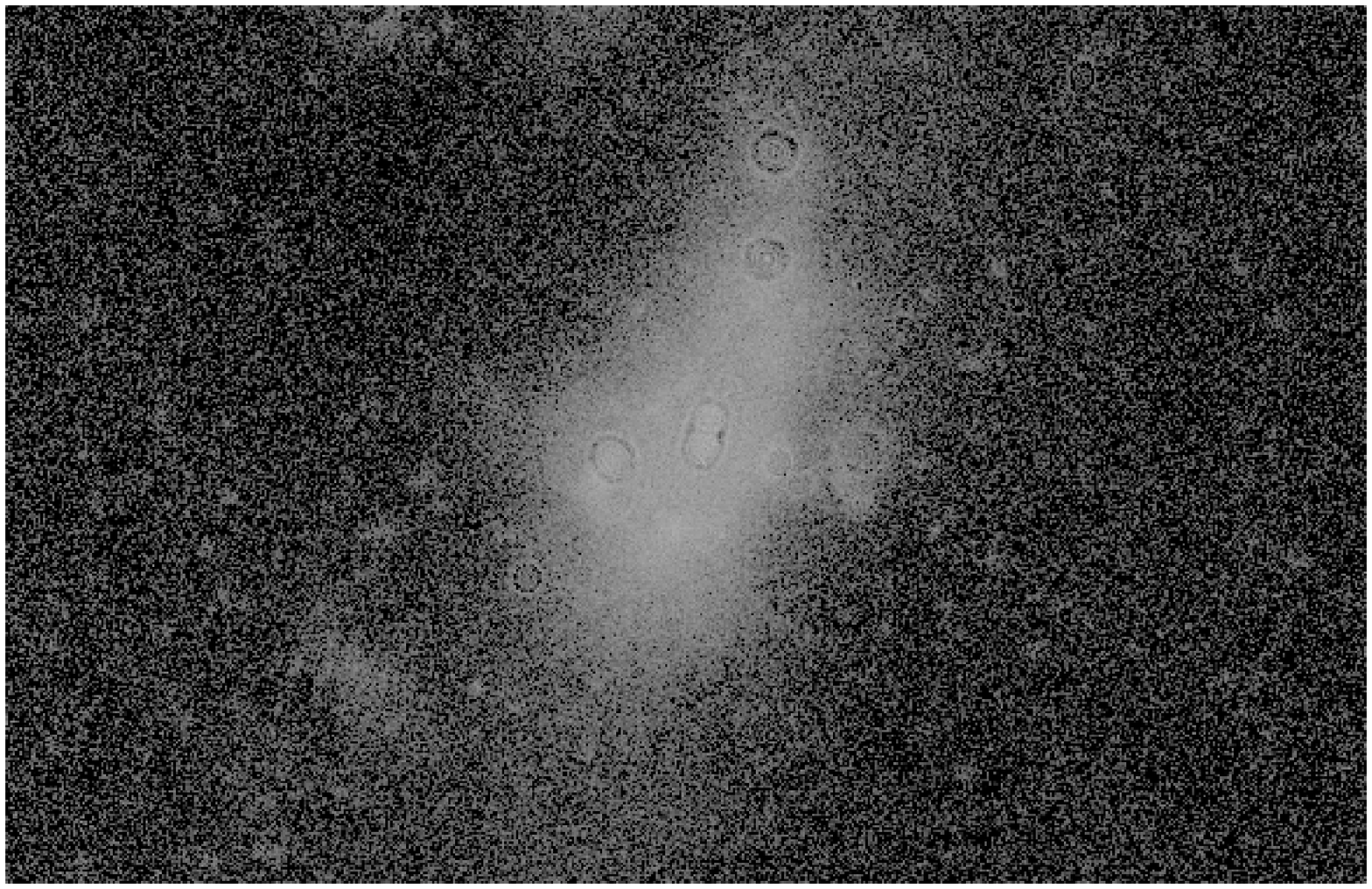}}
  \caption{(a, top left) Original I-band image of the central part of the cluster.  (b, top right) Image after subtraction of the BCG+ICL component using the model in (c). (c, bottom left)  {\tt IRAF ellipse} model. (d, bottom right) OVWAV reconstruction of the ICL in the same band.}
  \label{ignacio}
\end{figure*}

Figure~\ref{cen}  illustrates the well known challenges of detecting and characterizing the ICL in clusters: using photometry alone it is virtually impossible to separate the ICL from the halo (in our case haloes) of the BCG.  It is therefore customary to treat both components together \citep[e.g.][]{Gonzalez2005}.  We used the task {\tt Ellipse} within IRAF to build the surface brightness profiles of the BCG+ICL using as input our deep images together with a mask with all pixels brighter than $\rm{R}=25~\mathrm{mag~arcsec^{-2}}$ and all pixels within 15 pixels of the masked ones. The parameters of the task {\tt Ellipse} are used by the IRAF task {\tt Bmodel} to build a model image of the object and to generate the surface brightness profiles. To illustrate how well this procedure works, even in the densely populated central region of the cluster, we show in Figure~\ref{ignacio} our I-band image of the cluster compared with the same image subtracting the BCG+ICL component reconstructed using the IRAF procedures (top panels). 

We also experimented with the tools {\tt GALFIT} \cite{Peng} and {\tt OVWAV} \cite{DaRocha2005} that in principle should provide even more accurate models of the extended components. With GALFIT, which is probably the most powerful  publicly available tool to model galaxies (elliptical galaxies in particular), we were not able to obtain good models for the central galaxies either by fitting one galaxy at a time and masking the other two, or fitting the 3 galaxies together. Both procedures left large spourius residuals in the model subtracted image. OVWAV, that uses a wavelet analysis to detect and remove contaminating objects,  should allow to better model the extended components after cleaning. However, while we were able to obtain good OVWAV fits for our R-band FORS2 and I-band NTT images,  the procedure failed to converge for our V-band NTT image, in spite of extensive consultations with the author of the package.  Nevertheless, as shown in the bottom panels of Figure~\ref{ignacio}, the {\tt IRAF} reconstruction and the OVWAV procedure give remarkably consistent results, which make us confident that our method to model and characterize the extended BCG+ICL component is reliable and robust.  Figure~\ref{profis} presents the surface brightness profiles derived from the reconstructed images indicating that we reliably detect the ICL of the cluster out to radial distances of at least $r=350$ Kpc  from the cluster centre.

\begin{figure}
  \includegraphics[width=8.6cm]{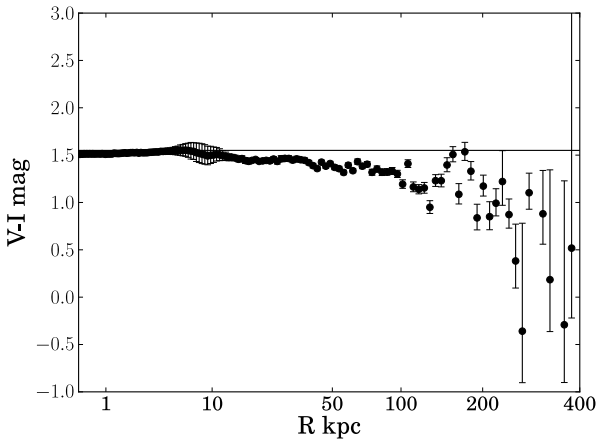}
  \caption{(V-I) colour of the BCG+ICL component as a function of radial distance. Special care was taken to match the fits of the two bands to avoid spurious differences due to misalignment and rebinning. Thus, the photometric error-bars provide accurate indications of the reality of the various features visible in this plot.}
  \label{cocor}
\end{figure}

In order to measure the colours of the BCG+ICL component we used the structural parameters for the fitting ellipses (ellipticity and orientation)  defined by the V-band to get the I-band profile brightness. This allow us to avoid false structures in the colour profiles due to mismatching of the ellipses, or to rebinning errors. The resulting colour profile presented in Figure~\ref{cocor} shows a gradient that can be traced out to at least 350~kpc, despite some fluctuations in the region outside of 150~kpc. The main body of the BCG extends out to approximately 10~kpc (see Figure~\ref{cen}), and then we see a bluer section with approximately no variations in colour extending from 10 to 40 kpc. From this point on we detect a distinct and continuous bluing out to approximately 350~kpc. This last section is probably a pure ICL component while the inner area probably still has a significant component of BCG halo.  A similar  trend has recently been observed in the extended halo of M87 by \cite{Rudick2010}, which they use to place constraints on the likely origin of the ICL in Virgo. We will return to this point in the analysis of our results.

\subsection{Spectroscopic properties of the ICL}
\label{sec:icl_spec}

\subsubsection{ Northern semi-mayor axis: ICL-N }

\begin{figure}
  \includegraphics[width=8.6cm]{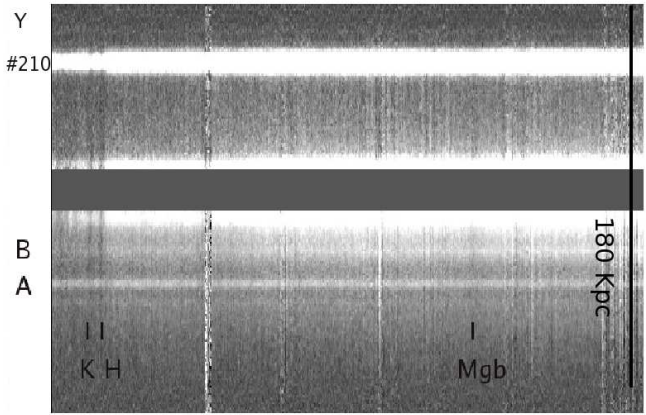}
  \caption{2-D spectrum of the ICL on the major axis from 4900\AA\ to 7550\AA. North is up. The spectra of galaxies A and \#210 in Figure~\ref{slits}, which are used for the spatial alignment of individual spectra, are identified. The object marked B appears as a clump in the ICL (Figure~\ref{slits}). It is in fact a background galaxy at z = 0.981. The main absorption lines of the ICL,  K, H, and Mgb, are indicated.}
  \label{ICLS}
\end{figure}

The combined 2-D spectrum of the major axis (N-S) component of the ICL and BGC halo is shown in Figure~\ref{ICLS}.

\begin{figure}
  \centering
  \vspace*{0.5cm}
  \includegraphics[width=8cm]{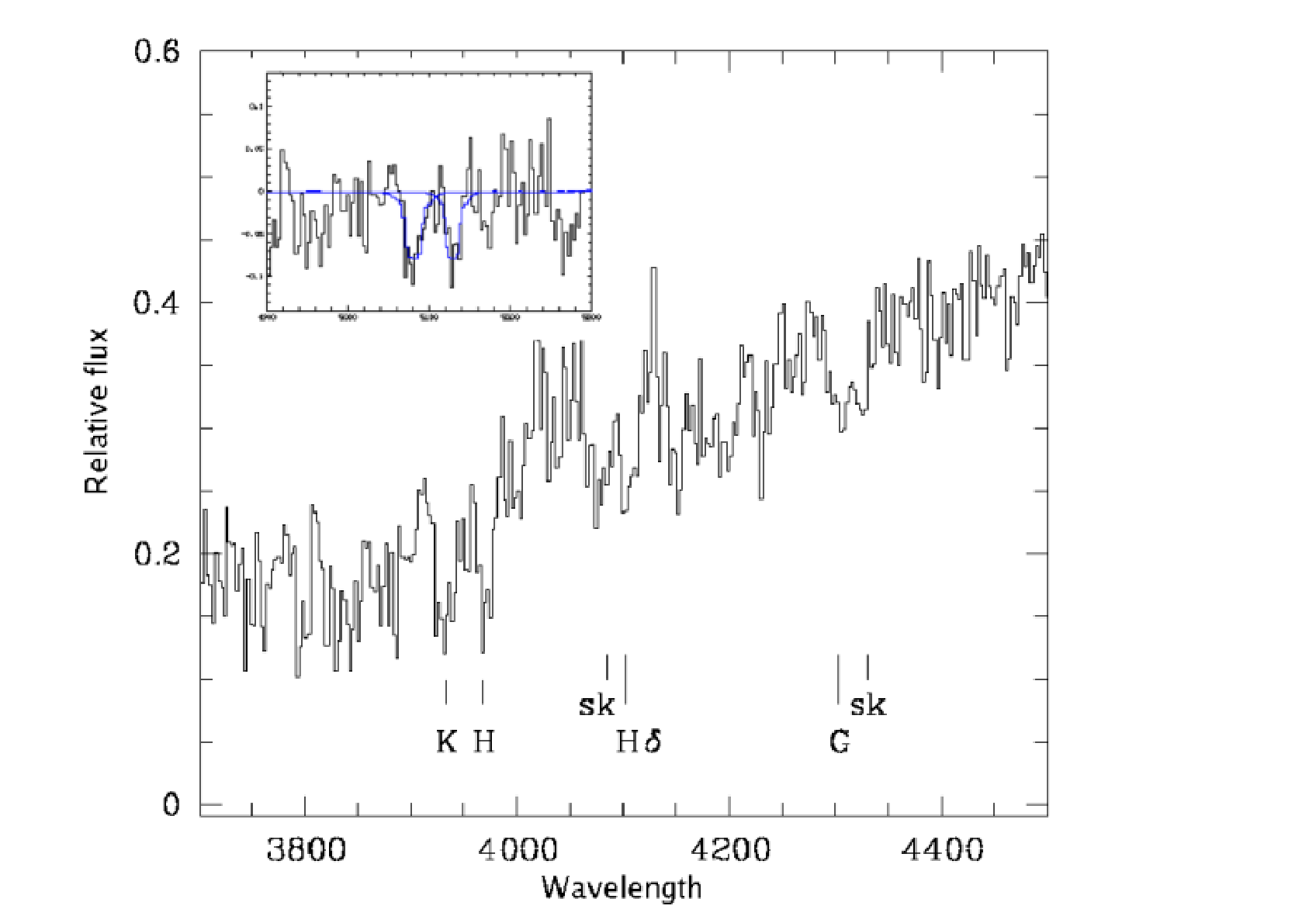}
  \vspace*{0.5cm}
  \includegraphics[width=8cm]{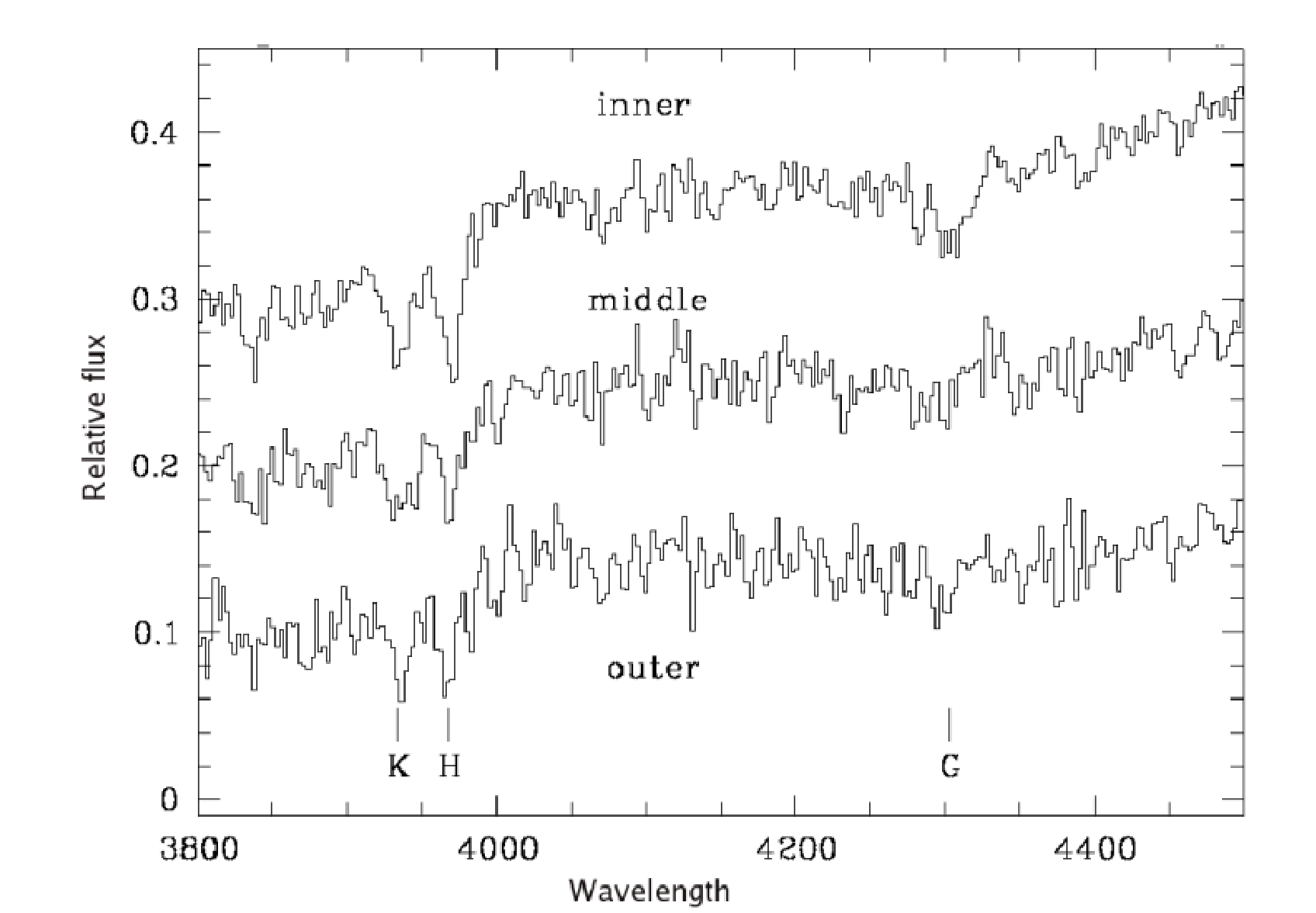}
  \vspace*{-.0cm}
  \caption{(a, Top). Spectrum of ICL-N rebinned to $z=0$ in the region of the 4000~\AA\ break  obtained by adding 22 rows in the region marked Z in Figure~\ref{slits}.    The insert shows the Gaussian fits used to estimate the velocity dispersions as discussed in the text.  (b, Bottom). Spectra of the ICL-S components.}
  \label{ICLNS}
\end{figure}

The sky background for the northern component (ICL-N) was measured in the slitlet marked Y in  Figure~\ref{slits}. After subtracting the sky from each of the 18 exposures of  ICL-N, and combining these 2-D spectra, we obtained a 1-D spectrum  by adding 22 rows at location Z in Figure~\ref{slits}.  The 1-D spectrum in the region of the 4000\AA\ break, rebinned to $z=0$, is shown in Figure~\ref{ICLNS}a. The radial velocities and 4000\AA\ break amplitude $\rm D(4000)$  measured in this spectrum are listed in Table~\ref{width} together with the values for other relevant regions of the cluster (see below).  We find a D(4000) step for ICL-N that is significantly shallower than that of the central cluster galaxies and of galaxy \#210, and is in fact typical of post-starburst (E+A) and of bulge-dominated spiral galaxies as discussed in Paper~I.  The lower value of D(4000) in this region is consistent with the bluer colour of the ICL seen in Figure~\ref{cocor}.  There is no evidence of  [OII] emission at the redshift of the cluster. 

The velocity dispersion of the H \& K lines of the ICL was measured as follows: a smoothed step function with D(4000) as step was used to model the continuum as shown in Figure~\ref{inter}a. The instrumental profile was deduced from the 5575\AA\  night-sky line, scaled to the depth of the absorption lines to be measured. It was assumed that the ICL line-profiles are Gaussian so  the observed widths are a convolution of the slit function with a Gaussian of  the same height. The line-widths were derived by minimizing the residuals after removing the continuum. The fit to H \& K in the ICL-N spectrum after continuum subtraction is shown in  the insert of Figure~\ref{ICLNS}a.

\subsubsection{Southern semi-major axis: ICL-S}

The sky background was measured at the southern end of the same slit used to measure the ICL  (the region below the absorption line symbols K, H, G  in Figure~\ref{ICLS}).  Three 1-D spectra were extracted at different locations of the ICL-S: an inner spectrum combining 8 rows located between the  spectrum of the southern component of the S-shaped arc and object B;  an middle spectrum combining 10 rows between galaxies A and B; and an outer spectrum obtained by combining 13 rows south of object A. The three resulting spectra are shown in Figure~\ref{ICLNS}b.  We observe a velocity gradient of $ \triangle V\sim180 \rm~ km~s^{-1}$ between the outer and inner positions, and a corresponding gradient in D(4000) that is consistent with the bluer colour of the ICL relative to the central galaxies. 

\subsubsection{ ICL minor axis}
\label{ICLminor}

\begin{figure}
   \centering
   \vspace*{0.5cm}
   \includegraphics[width=8cm]{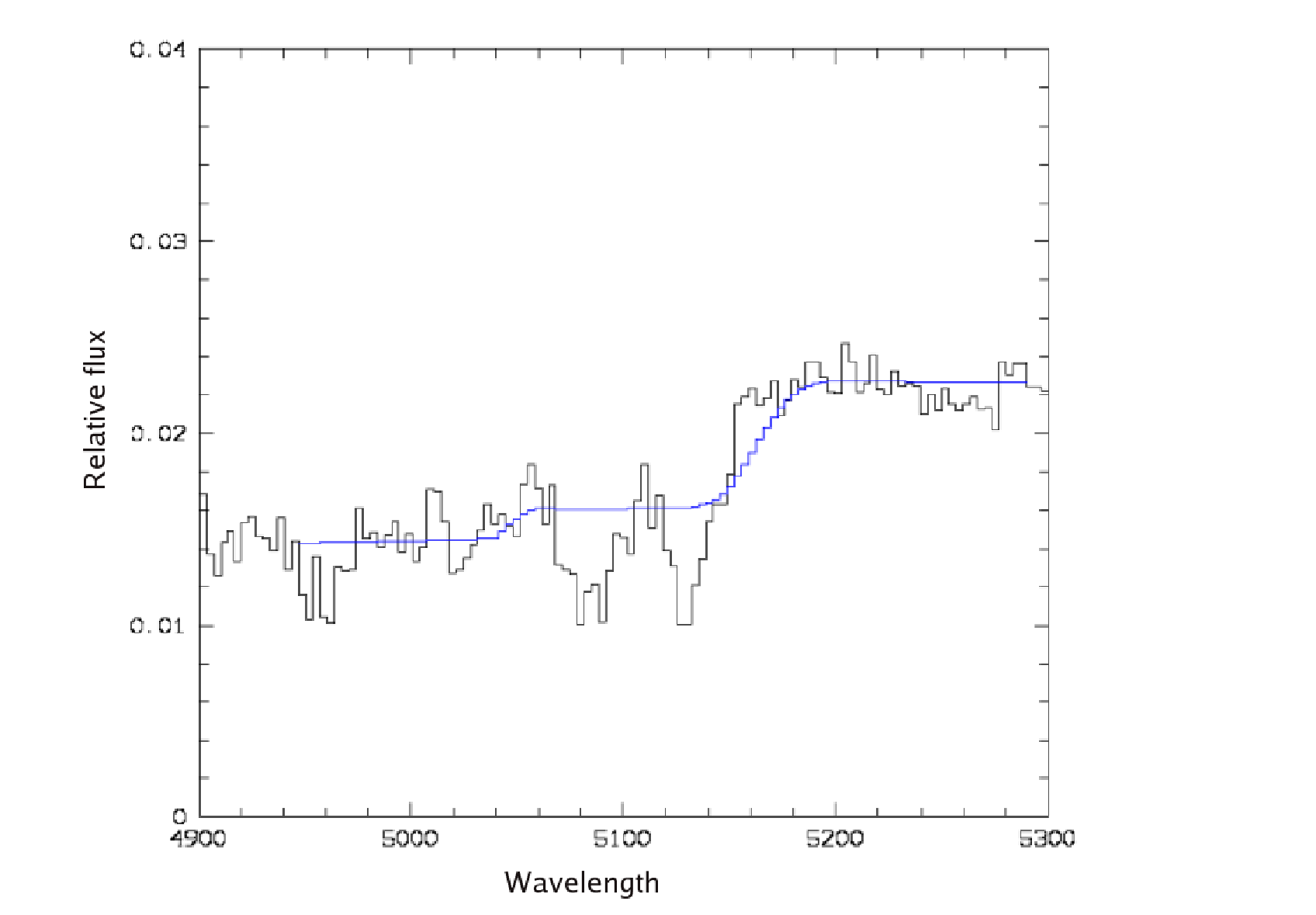}
   \vspace*{0.5cm}
   \includegraphics[width=8cm]{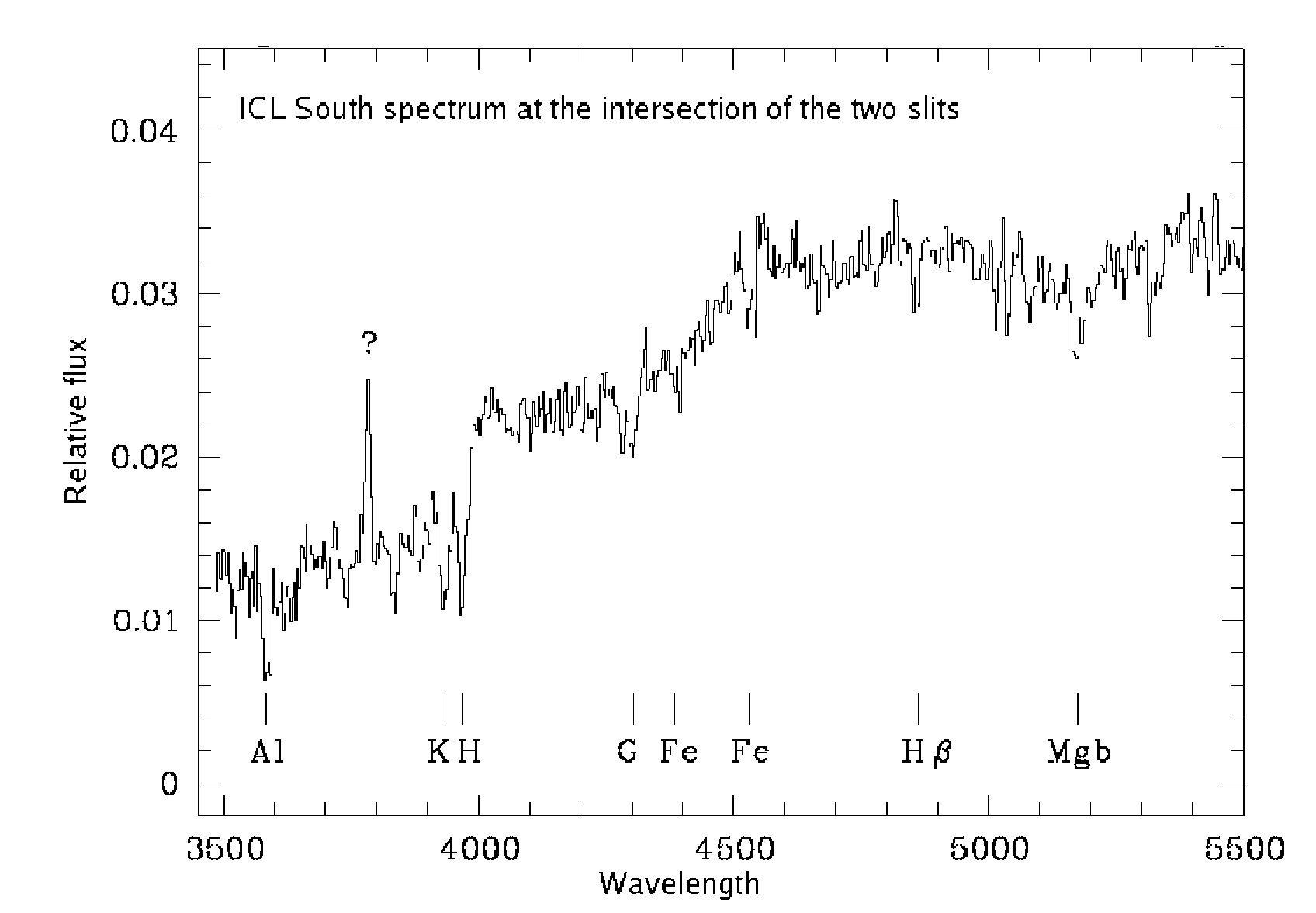}
   \vspace*{-.0cm}
   \caption{(a, top) Spectrum of the ICL-S at the intersection of the two slits with a model continuum superimposed in blue. (b, bottom) Same spectrum corrected for redshift where the metallic lines are indicated. The feature marked with a question mark could not be identified with any nebular emission-line at the redshift of the cluster, nor do we find a secondary absorption line system in the spectrum of the ICL component.}
   \label{inter}
\end{figure}

The sky background was measured on both sides of the ICL as indicated in Figure~\ref{slits}. We obtained three 1-D spectra by combining 13 rows starting from south of galaxy \#493 outward; 16 rows on the opposite side; and 10 rows in the inner region. We find a large E-W radial velocity gradient of $\triangle V\sim 300\rm~ km~s^{-1}$ and, consistently with ICL-S, D(4000) is significantly smaller in the outer regions than in the middle.  We combined the spectra in the region where the slits of the major and minor axes cross. This gives a reasonably high S/N spectrum (Figure~\ref{inter}b) where we clearly detect  strong metallic absorption lines - Al III 3584\AA, Fe 4383\AA,  and 4531\AA. The $\rm H\delta$ absorption line in this region is weaker than in ICL-N. 

\subsubsection{Kinematics of the ICL}

\begin{figure}\centering\vspace*{0.5cm}
  \includegraphics[width=8.6cm]{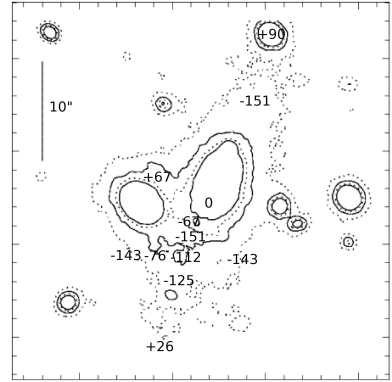}
  \caption{Line-of-sight radial velocities relative to galaxy \#509, illustrating the caothic motion of the diffuse component, as discussed in the text.}
  \label{velocity}
\end{figure}

Table~\ref{width} and Figure~\ref{velocity} summarize the velocity information for  the various components of the ICL discussed above. The radial velocities,  $\rm \triangle V~km~s^{-1}$,  are in the rest frame of galaxy \#509, which is at $z = 0.2929$.   The table shows that the radial velocities of the ICL are quite chaotic, and that, within the rather large errors, the velocity dispersions seem to dominate over velocity differences.  There is clear indication of a systematic increase in velocity dispersion with distance to the cluster centre,  which can be interpreted as evidence for a true ICL component dynamically decoupled from the BCGs, more representative of the cluster as a whole. The least uncertain velocity dispersion measurements are for the intersection of the NS and EW stilts,  ICL-N, and ICL-S (middle) that have the lowest formal errors.  According to our photometry, only ICL-S (middle) would be a bona-fide ICL component, while ICL-N and the intersection region are well within the BCG haloes. Taking the average of ICL-N, ICL-S(inner), and the slit intersection region as representative of the BCG haloes, we get a halo velocity dispersion of  $414\pm114~km~s^{-1}$. We do not have a stellar template to calibrate the instrumental width of a single (K-type) star, so we used galaxy \#210  that was observed with a reasonably good S/N to infer a lower limit for the true velocity dispersions. We thus get $\sigma_{BCG-halo}\sim 353\pm135 km~s^{-1}$ for the BCG haloes, and $\sigma_{ICL}\sim 478\pm113\rm~km~s^{-1}$ for the ICL (that is ICL-S(middle)). This is consistent, within the errors, with both values for the velocity dispersion of the cluster given in Table~\ref{props}.  As expected, the velocity dispersion of the ICL formally larger than that of the BCG halo, but the difference is not significant given the large errors.

\begin{table*}
\caption{Integrated Properties of the ICL of the cluster RX J0054.0-2823}                   
\centering                                    
\begin{tabular}{ l c c c c c}          
\hline\hline                        
Location   & $\triangle \alpha$ (sec)   &   $\rm \triangle \delta $ ($\arcsec$) & $ \triangle V $ (km/s)  & $\sigma_V$ (km/s) & D(4000) \\     
\hline                                    
Center (galaxy $\#509$)  	&   0              	&      0          	&  --                    		&   --                      		& --              		 \\      
Galaxy $\#210$           	&  +0.36           	&     +3.5        	& $+93 \pm 30$       	& $216  \pm 34$            	&  $1.83 \pm 0.01$ 	\\
ICL N       		 	&  +0.24           	&     +8.6        	& $-151 \pm 43$       	& $436 \pm 102$            	&  $1.39 \pm 0.02$	 \\
ICL-S (inner)            	&  -0.17           	&     -4.4        	& $-151 \pm 30$       	& $318  \pm 135$            	&  $1.55 \pm 0.01$ 	\\
ICL-S (middle)          	&  -0.25           	&     -7.5        	& $-125 \pm 43$       	& $525 \pm 102$            	&  $1.51 \pm 0.02$ 	\\
ICL-S (outer)            	&  -0.35           	&     -11.1       	& $+26 \pm 0.30$     & --                        		&  $1.39 \pm 0.02$ 	\\
ICL Minor axis (east)    	&  -0.62           	&     -4.8        	& $+143 \pm 43$      & $550   \pm 148$            	&  $1.46 \pm 0.02$ 	\\
ICL Minor axis (middle)  	&  -0.29           	&     -4.9        	& $-76 \pm 43$       	& $550  \pm 148$            	&  $1.54 \pm 0.01$ 	\\
ICL Minor axis (west)    	&  +0.12           	&     -5.4        	& $-143 \pm 43$       	& $593   \pm 161$            	&  $1.44 \pm 0.02$ 	\\
Intersection of slits    	&  -0.19           	&     -5.1        	& $-112 \pm 30$       	& $487 \pm 76$            	&  $1.54 \pm 0.01$ 	\\
Object B                 	&  -0.22           	&     -6.2        	&  --                    		&  --                      		&   --             \\
Object A                 	&  -0.29           	&     -9.1	      	&  --                    		&  --                       		&   --             \\
Arc North                	&  -0.35           	&     +1.32       	& $+67 \pm 30$       	& $335  \pm 102$            	&  $1.67 \pm 0.01$ 	\\
Arc South 	         	&  -0.11           	&     -3.1        	& $-67 \pm 30$       	& $425  \pm 135$            	&  $1.59 \pm 0.01$ 	\\

\hline
\end{tabular}
\label{width}
\end{table*}

\subsubsection{The S-shaped arc}

\begin{figure}
  \vspace*{0.75cm}
  \includegraphics[width=8.6cm]{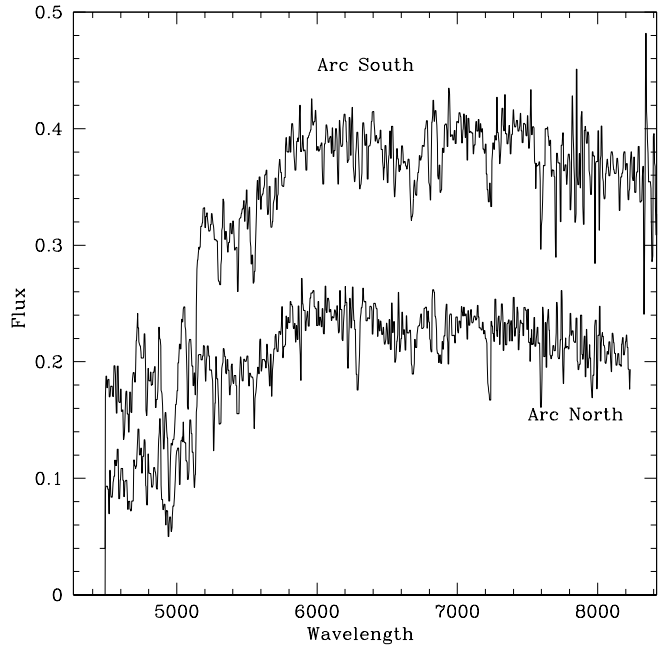}
  \caption{ Extracted spectra of the two components of the S-shaped arc. No emission lines are visible on either component, which otherwise have spectra typical of faint absorption-line galaxies in the cluster.}
  \label{zARC}
\end{figure}

Figure~\ref{zARC} shows 1-D extractions of the North and the South components of the S-shaped arc. No emission lines are visible in either component, which otherwise have spectra typical of faint absorption-line galaxies in the cluster (Paper~I). The two spectra, however, are not identical and in fact have slightly different redshifts, although the difference of $\rm 165~ km~s^{-1}$ is only at the 2$\sigma$ level. Both components show clear Balmer lines in absorption, but those of the southern component are stronger.  We find no evidence, therefore, that either of the components of the S-shaped arc could be the gravitationally lensed image of a background galaxy. 

%
%

\section{Discussion}
\label{SECanal}

\subsection{Dynamics of the cluster and origins of the ICL}


The fact that the cluster has rather regular X-ray contours indicates a reasonably relaxed centrally symmetric gravitational potential. It seems plausible, therefore, to assume that the three central E-galaxies are in fact close neighbors. The likely outcome of this configuration is that the dumbbell will circularize its orbit and that eventually the three galaxies will merge \citep{Brough2008,Pimbblet2008}. In the process, they create a strong varying gravitational tidal field that will shred to pieces any small galaxy that ventures to the central region of the cluster, particularly late-type galaxies. This could be the source of the two pieces of our S-shaped arc, that has colours and spectral types consistent with those of faint galaxies in the cluster, but very similar redshifts to the surrounding large ellipticals.  Thus, the three massive central elliptical galaxies could form a sort of efficient galaxy grinding machine, rapidly filling the central regions of the cluster with debris and dispersed stars. This is consistent with the complex pattern in the radial velocities of the diffuse components (Figure~\ref{velocity}) and their large velocity dispersions. 

This is also consistent with the clear outward bluing of the ICL, both in the photometry and in the spectra. Similar bluing trends have been found by \cite{Rudick2009} for Virgo, and by \cite{jee2010} for CL0024+17 at z=0.4, who argued that the bluing provided evidence of ongoing stripping of stars. If the bulk of the stars were stripped a few Gyrs earlier the ICL would be redder if the cluster galaxies  continue to form new stars \citep{sommerlarsen2005}. In fact, our spectra indicates that the ICL luminosity is dominated by star younger than 2.5 Gyrs. This explanation is also consistent with the colour gradient observed in the red-sequence of clusters (e.g. Figure~\ref{cmd}) if we posit that the main donors of ICL and BCG halo stars are faint galaxies, and with the measured values of D(4000). Our ``galaxy grinding machine" model would provide a mechanism to accomplish this efficiently, although we also believe that there might be other contributors to the observed bluing, e.g. unmasked faint galaxies, more of which are forming stars in the periphery of the cluster than near the centre.

\subsection{Baryons}

\begin{figure}
  \includegraphics[width=8.6cm]{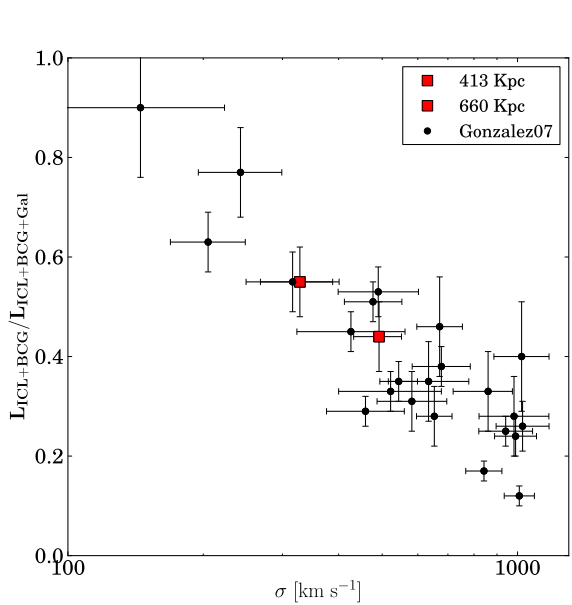}
  \caption{Fraction of the total light in clusters that is emitted by the BCG+ICL component as a function of velocity dispersion. The filled squares are the data from GZZ and the red square are the two values corresponding to the two possible values of the velocity dispersion of our cluster listed in Table \ref{props}. For both values our cluster fits the correlation rather well.}
  \label{gonza}
\end{figure}

GZZ presented a comprehensive study of the ICL component in 23 clusters from the \cite{Gonzalez2005} sample of clusters with well determined dynamical masses and X-ray luminosities. The central aim of GZZ was to account for all baryons in clusters, and the inclusion of the ICL, which may contain as much as 50\% of all the baryonic mass, was of paramount importance. In fact, because of the problems to disentangle the ICL from the halo of the BCG mentioned above, GZZ actually measured the fraction contained by both components together that for the purpose of counting baryons is  a valid approach.  GZZ found a correlation between the velocity dispersion of their clusters and the fraction of the total cluster light that is contained by the BCG+ICL component. Figure~\ref{gonza} reproduces the GZZ correlation using the values tabulated in their paper. The BCG+ICL component dominates the stellar mass for small mass clusters, but is less than 40\% for the most massive ones.  GZZ argued that there was a good physical basis to expect such correlation and ruled out the possibility that the correlation results from selection biases in their data, in particular the radial dependence of the BCG+ICL component, which is more centrally concentrated than the galaxies.  Recently, however, \cite{Puchwein2010} failed to reproduce such a strong correlation using state of the art hydrodynamical simulations, although they did not consider low-mass clusters that are actually the ones that define the steep correlation of GZZ. The slope of the models is roughly in agreement with the observations for clusters more massive than  $\sim10^{13}M_{\sun}$, but in this mass range the models still predict about two times more BCG+ICL light than is observed.   \cite{Puchwein2010} argued that they did not expect significant changes in their estimates from improvements in the simulations, and therefore shifted the burden of proof on improved observations. 

\begin{figure}
  \includegraphics[width=8.6cm]{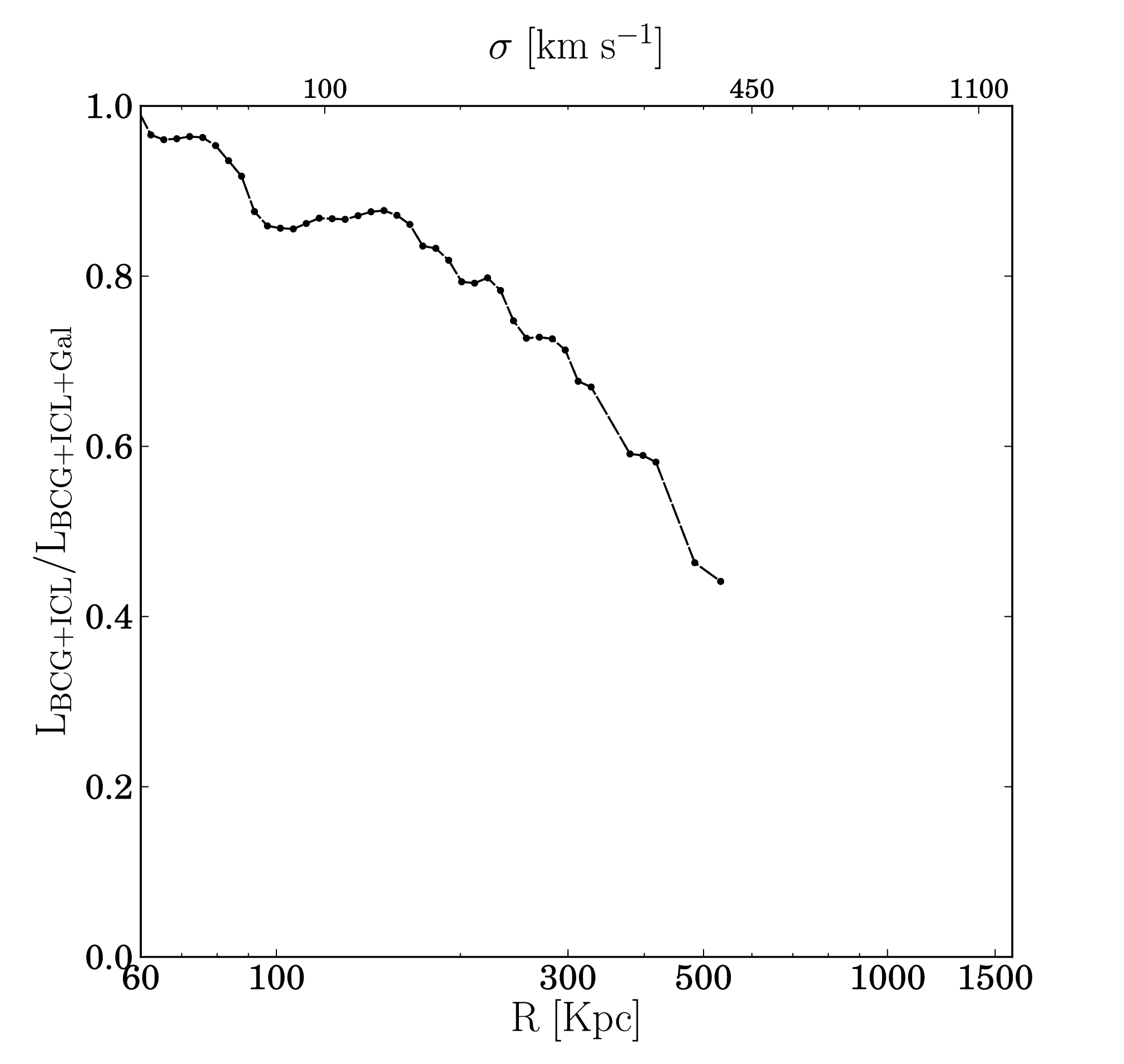}
  \caption{Fraction of the total light in the cluster contained in the BCG+ICL component as function of the enclosing radius (bottom axis). The top axis scale is the equivalent velocity dispersion using the relation $\sigma-r_{500}$ described in the text and can be used to compare with figure~\ref{gonza}. Note the significant radial dependence of the measured fraction. We don't show the fraction with radius larger than $r=600~Kpc$ because we don't detect the ICL component beyond that radius.}
  \label{totprof}
\end{figure}

The BCG+ICL fraction for our cluster is shown by the two coloured open squares in Figure~\ref{gonza}, corresponding to the two possible values for the velocity dispersion of the cluster. Both points are seen to fit very well in the GZZ correlation. This happens because, as indicated by GZZ, and shown in Figure~\ref{totprof} for our cluster, the BCG+ICL fraction decreases with increasing radius and we use $R_{500}$ that scales with velocity dispersion to calculate the fractions.  In fact, our cluster fits the GZZ relation for any value of $\sigma$ in the range allowed by the observations ($330-500\rm~km~s^{-1}$).  This suggests that the steep part of the GZZ relation - the region of small cluster masses - may indeed be affected by the strong radial gradient in the BCG+ICL fraction (Figure~\ref{totprof}).  The errors that propagate to the BCG+ICL fractions through the measurement uncertainties in the velocity dispersions of low-mass clusters via the $\sigma-R_{500}$ relation, and the strong radial dependence of the BCG+ICL fraction, actually  blur the low-mass end of the correlation. In fact, if clusters with velocity dispersions lower than $500\rm~km~s^{-1}$ are removed from the sample, the remaining (massive) clusters define a trend that is consistent with the trend predicted by the numerical simulations of Puchwein et al., although the discrepancy with the absolute values of the BCG+ICL fractions remains.

We suspect that the discrepancy may be related to the way in which colours and magnitudes are assigned to the numerical stars as a function of mass, age, and chemical composition, and their assumed IMF. Unfortunately, however, \cite{Puchwein2010} and the references they cite do not provide enough details about their treatment of stellar evolution to look deeper into the models.

\section{Summary and Conclusions}
\label{SECconcl}
 
We have presented deep photometry and spectroscopy of the diffuse intracluster light (ICL) of an intermediate redshift cluster at z=0.3. We show that despite the redshift we are able to measure both the photometric and kinematic properties of the ICL out to distances comparable to the distances reached by state-of-the-art observations of local clusters such as Virgo. This is because, being more compact, we have less problems to subtract the sky background and the contamination by  large scale features like Galactic cirrus. And of course we are able to observe with 8m-class telescopes.  We found that globally the BCG+ICL component is reasonably well characterized by a de Vaucouleurs $R^{1/4}$ law, and we detect a significant outward bluing of the diffuse component. At a distance between 40~kpc and 350~kpc from the cluster centre, the colour gradient flattens and the slope of the surface brightness profile steepens probably signaling the region of pure ICL light.  We also find that the D(4000) index, which is an indicator of stellar populations, is shallower in these outer regions   and consistent with the value observed for faint galaxies in the cluster. This, combined by the rather chaotic kinematics of the ICL at large radii, is consistent with the observations in lower-z clusters of filamentary structures in the ICL. Our cluster is characterized by a triplet of bright elliptical galaxies in its centre, two of which form a close dumbbell pair. We argue that this structure will eventually merge thus significantly increasing the BCG+ICL component. In spite of its dynamical youth, however, the fraction of BCG+ICL  to total light in our cluster is consistent with the values observed for similar mass clusters at low redshifts. If real, this indicates that most of the ICL must be formed early in the merging history of clusters.  However, the radial velocity histogram of our cluster is consistent with two different values for the velocity dispersion, and curiously the cluster fits the correlation between BCG+ICL fraction and velocity dispersion for both velocity dispersions equally well. This happens because there is a well known relationship between BCG+ICL fraction and radius, and we integrate the light out to a radius that scales with velocity dispersion. This leads us to suggest that the correlation between BCG+ICL and velocity dispersion may be much weaker than previously thought, which is consistent with the predictions of hydrodynamical numerical simulations. However, the simulations predict BCG+ICL fractions on average a factor of $\sim2$ larger than observed (with BCG+ICL fractions up to $70\%$ and with the ICL alone having  $45\%$ of the stellar mass), but unfortunately the theoretical papers do not provide sufficient details about the modeling to allow a better understanding of the discrepancy. In particular, it is not clear how these simulations model stellar evolution, nor whether the models actually reproduce the observed colours and D(4000) steps of the ICL.  We are confident in our measurement of the BCG+ICL fraction in RX~J0054.0-2823 and therefore that the discrepancy does not lie with the observations.   

The main conclusions of our paper can be summarized as follows:

\begin{enumerate}
\item The S-shaped arc in the ROSAT cluster RX~J0054.0-2823 is not the gravitationally lensed image of a background galaxy, but the remnant  of two faint cluster galaxies in the act of being tidally destroyed by the gravitational field of the three giant elliptical galaxies in the cluster centre;

\item In spite of the relatively large redshift of the cluster ($z\sim0.3$) we were able to clearly detect its diffuse intracluster-light (ICL) component and to measure its broad-band colours and spectral signatures. The bluer colour and shallower spectral indexes of the ICL, and its chaotic motions are consistent with the idea that the ICL originates from tidally disrupted galaxies. 

\item For the ICL we measured D(4000)=1.39$\pm$0.02, typical of post-starburst (E+A) galaxies. This  value is significantly different from D(4000)=1.83$\pm$0.01 measured for the central galaxies, and is consistent with the bluer colour of the ICL;

\item The bi-modal radial velocity distribution of the cluster galaxies makes its velocity dispersion rather uncertain, despite the fact that we have velocities for close to 100 cluster galaxies.  However, we find that the cluster fits the correlation between the fraction of the light in the BCG+ICL component and velocity dispersion found by GZZ for any value of the velocity dispersion in the range allowed by the observations. This led us to question the correlation given the rather large observational uncertainties in the velocity dispersions of the low-mass clusters in the GZZ sample. However, our cluster shows the same discrepancy between the observed BCG+ICL fraction and the values predicted by numerical simulations.

\item The BCG+ICL fraction of our cluster is undistinguishable from that of similar mass clusters at lower redshift in the  GZZ sample. This indicates that most of the diffuse component in clusters is already in place at z=0.3. 

\end{enumerate}

\section*{Acknowledgments}

This work was based on observations obtained in service mode at the European Southern Observatory at Paranal. I. Toledo would like to thank ESO for the hospitality at the Vitacura Offices throughout his work. H. Quintana thanks partial support from FONDAP “Centro de Astrof´ısica”. P. Zelaya acknowledges a studentship from CONICYT. We thanks C. De Rocha for assisting as with the OVWav algorithm and C. Peng for giving us access to GALFIT 3.0 .

\bsp
\label{lastpage}

\end{document}